%% file: main.tex
\documentclass{ieeeaccess}

\usepackage{amsmath}

\usepackage{algorithm,algorithmic}

\usepackage[utf8]{inputenc}
\usepackage[english]{babel}
\usepackage{tikz}
\usetikzlibrary{spy}
\usepackage[export]{adjustbox}
\usepackage{transparent}
\usepackage{amssymb}
\usepackage{bbold}
\usepackage{todonotes}
\setuptodonotes{inline}
\usepackage{cite}
\usepackage{amssymb}
\usepackage{textcomp}
\usepackage{url}
\def\BibTeX{{\rm B\kern-.05em{\sc i\kern-.025em b}\kern-.08em
    T\kern-.1667em\lower.7ex\hbox{E}\kern-.125emX}}
    
\NewSpotColorSpace{PANTONE}
\AddSpotColor{PANTONE} {PANTONE3015C} {PANTONE\SpotSpace 3015\SpotSpace C} {1 0.3 0 0.2}
\SetPageColorSpace{PANTONE}%

\newcommand{\rigid}{\texttt{rigid}}
\newcommand{\shear}{\texttt{shear}}
\newcommand{\nonlinear}{\texttt{nonlinear}}

\newcommand{\R}{\mathbb{R}}
\newcommand{\N}{\mathbb{N}}
\renewcommand{\d}{\mathrm{d}}

\newcommand{\imspace}{\mathbb{U}}

\newcommand{\vecspace}{\mathbb{V}}

\newcommand{\RegIm}{\mathcal{R}}
\newcommand{\RegVf}{\mathcal{S}}
\newcommand{\ParamIm}{\alpha}
\newcommand{\coupling}{\mathcal{H}}
\newcommand{\fid}{\mathcal{D}}

\newcommand{\fwd}{A}
\newcommand{\interp}{J}
\newcommand{\data}{f}
\newcommand{\vf}{\phi}
\newcommand{\pvf}{\varphi}
\newcommand{\sinfo}{v}

\newcommand{\norm}[1]{\left\|#1\right\|}
\newcommand{\argmin}{\operatorname{arg}\min}
\newcommand{\argmax}{\operatorname{arg}\max}
\newcommand{\tv}{\operatorname{TV}}
\newcommand{\dtv}{\operatorname{dTV}}

\newcommand{\prox}{\operatorname{prox}}
\newcommand{\id}{\operatorname{id}}

\newcommand{\LB}[1]{{\color{red}#1}}

\renewcommand{\LB}[1]{#1}

\newcommand{\revise}[1]{{\color{olive}#1}}
\renewcommand{\revise}[1]{#1}

\def\PicWidth{3.5cm}%
\def\ZoomWidth{2.3cm}%
\def\MagnifyingFactor{3}%

\newcommand{\PlotImage}[1]{{\includegraphics[width=\PicWidth]{#1}}}

\newcommand{\PlotFigZoomSpy}[5]{%
\begin{tikzpicture}[spy using outlines, x=\PicWidth, y=\PicWidth, inner sep=0pt]
\draw (0, 0) node [anchor=south west] {{\PlotImage{#3}}};%
\spy [color=#5, circle, draw, width=\ZoomWidth, height=\ZoomWidth, magnification=\MagnifyingFactor,
connect spies] on #1 in node [left] at #2;
\draw (.02, .02) node [anchor=south west] {\parbox{\PicWidth}{\color{white}\footnotesize #4\phantom{fg}}};%
\end{tikzpicture}}%

\newcommand{\PlotFigZoomSpyMetrics}[7]{%
\begin{tikzpicture}[spy using outlines, x=\PicWidth, y=\PicWidth, inner sep=0pt]
\draw (0, 0) node [anchor=south west] {{\PlotImage{#3}}};%
\spy [color=#7, circle, draw, width=\ZoomWidth, height=\ZoomWidth, magnification=\MagnifyingFactor,
connect spies] on #1 in node [left] at #2;
\draw (.02, .01) node [anchor=south west] {\parbox{1.5cm}{\color{white}\footnotesize #4\phantom{fg}}};%
\draw (0.45, 0.0) node [anchor=south west] {\parbox{1.5cm}{\color{white}\footnotesize 
\begin{tabular}{lr}
     SSIM & #5\\
     RD & #6
\end{tabular}
}};%
\end{tikzpicture}}%

\newcommand{\PlotFig}[2]{%
\begin{tikzpicture}[x=\PicWidth, y=\PicWidth, inner sep=0pt]%
\draw (0, 0) node [anchor=south west] {{\PlotImage{#1}}};%
\draw (.02, .02) node [anchor=south west] {\parbox{\PicWidth}{\color{white}\footnotesize #2\phantom{fg}}};%
\end{tikzpicture}}%

\newcommand{\PlotFigs}[2]{%
\begin{tikzpicture}[spy using outlines, inner sep=0pt]%
\draw (0cm, 0cm) node [anchor=south west] {#1};%
\draw (0cm, -.1cm) node [anchor=north west] {\footnotesize #2};%
\end{tikzpicture}}%

\begin{document}
\doi{10.1109/ACCESS.2020.3043638}

\title{Robust Image Reconstruction with Misaligned Structural Information}
\author{\uppercase{Leon Bungert}\authorrefmark{1}, 
\uppercase{Matthias J. Ehrhardt\authorrefmark{2}}}
\address[1]{
Department of Mathematics, University of Erlangen, Germany (e-mail: leon@bungert@fau.de)}
\address[2]{
Institute for Mathematical Innovation, University of Bath, UK (e-mail: m.ehrhardt@bath.ac.uk)}
\tfootnote{LB was supported by the European Union’s Horizon 2020 research and innovation programme under the Marie Sk\l{}odowska-Curie grant agreement No 777826 (NoMADS). MJE acknowledges support from the EPSRC (EP/S026045/1, EP/T026693/1), the
Faraday Institution (EP/T007745/1) and the Leverhulme Trust (ECF-2019-478).}

\markboth
{Bungert and Ehrhardt: Robust Image Reconstruction with Misaligned Structural Information}
{Bungert and Ehrhardt: Robust Image Reconstruction with Misaligned Structural Information}

\corresp{Corresponding author: Leon Bungert (e-mail: leon.bungert@fau.de).}

\begin{abstract}
Multi-modality (or multi-channel) imaging is becoming increasingly important and more widely available, e.g. hyperspectral imaging in remote sensing, spectral CT in material sciences as well as multi-contrast MRI and PET-MR in medicine. Research in the last decades resulted in a plethora of mathematical methods to combine data from several modalities. State-of-the-art methods, often formulated as variational regularization, have shown to significantly improve image reconstruction both quantitatively and qualitatively. Almost all of these models rely on the assumption that the modalities are perfectly registered, which is not the case in most real world applications. We propose a variational framework which jointly performs reconstruction and registration, thereby overcoming this hurdle. \LB{Our approach is the first to achieve this for different modalities and outranks established approaches in terms of accuracy of both reconstruction and registration.} \revise{Numerical results on simulated and real data show the potential of the proposed strategy for various applications in multi-contrast MRI, PET-MR, and hyperspectral imaging:} typical misalignments between modalities such as rotations, translations, zooms can be effectively corrected during the reconstruction process. Therefore the proposed framework allows the robust exploitation of shared information across multiple modalities under real conditions. 
\end{abstract}


\titlepgskip=-15pt

\maketitle

\section{Introduction}
\label{s:introduction}
\IEEEPARstart{T}{he aim} \revise{of this paper is to find an approximate solution to an ill-posed inverse problem of the form $Au=f$, within the framework of structural regularization.}
Here $\fwd$ denotes the forward operator and $f$ is measured data which is typically corrupted by noise. Examples include magnetic resonance imaging (MRI), positron emission tomography (PET), computed tomography (CT), image denoising, deblurring, or super-resolution, or possibly a combination of these tasks. A widely used approach of obtaining approximate solutions of inverse problems is \emph{variational regularization} where prior knowledge, like for instance sparsity of the solution or its gradient, is enforced through a regularization functional, see~\cite{Scherzer2008book, Ito2014book, Benning2018actanumerica} and references therein.

In multi-modality imaging one is often in possession of a specific piece of prior knowledge, namely a side information $\sinfo$ which is known to have some ``common structure'' with the true solution $u$ of the inverse problem \cite{Ehrhardt2020chapter}. The literature on this topic is rich with some works as early as the 1990's so we only list a couple of key papers here. For instance, $\sinfo$ could be a high-resolution photograph which assists the reconstruction of low-resolution hyperspectral images~\cite{Ballester2006, Moeller2012, Loncan2015, Duran2017, bungert2018blind} or an anatomical (MRI or CT) image for the reconstruction of PET images~\cite{Leahy1991, Gindi1993, Bowsher2004, ehrhardt2016pet, Mehranian2017petplusmri, rasch2017joint, Tsai2018mic, Tsai2019petct}, see Figure~\ref{fig:motivation}. This strategy has also been used for functional MRI (fMRI)~\cite{Rasch2018}, spectral CT \cite{Kazantsev2018}, electrical impedance tomography (EIT)~\cite{Kolehmainen2019eit} and multi-contrast MRI~\cite{Bilgic2011, Ehrhardt2016mri, Huang2014multicontrastmri, Song2020mri, Obert2020, Teh2020}.

While these approaches can greatly improve the image reconstruction, they all assume that the target solution $u$ and the side information $v$ are perfectly aligned. If they are badly aligned, most of these methods yield unsatisfactory reconstructions, see Figure~\ref{fig:misaligned} for an illustration. 
In real-world applications such misalignment typically cannot be avoided since frequently the acquisitions of $v$ and $f$ happen at different times and through different modalities, see e.g.~\cite{yokoya2017hyperspectral}. 

To mitigate this problem, we consider the following joint reconstruction and registration model
\revise{%
\begin{align}\label{eq:joint_prob}
\min_{u,\vf}\fid(\fwd (u\circ\vf);\data)+\RegIm(u;\sinfo)+\RegVf(\vf).
\end{align}}
This model seeks to reconstruct an image $u$ together with a deformation field $\vf$, such that $u$ has common structure with the side information $\sinfo$ (enforced through the regularization term $\RegIm(u;\sinfo)$) and the deformed image $u\circ\vf$ explains the data $f$ well \revise{(enforced through the data fidelity term $\fid(A(u\circ\vf); f)$)}. In addition, functional $\RegVf$ can be chosen to suitably regularize~$\vf$. Note that the minimization can be restricted to subsets of admissible images or deformation fields. In particular, one can use \emph{parametric deformations} in order to only allow for rigid, affine, diffeomorphic, and other deformations.

\LB{The main advantage of our model over the ones discussed in the next section is its jointness which allows for more accurate estimations of the misalignment and leads to better image quality.}
\input{figures/fig_1-2}
\subsection{Related work}
\label{ss:related_work}
A common strategy in the literature is to perform a three-step strategy: first reconstruct $u$ without the side information $v$, then to register $u$ and $v$ and finally to reconstruct again using a registered side information (cf.~\eqref{eq:alternating} below). This has been considered for PET-MR \cite{Ehrhardt2019pmb}. The registration of different modalities is often achieved through maximizing the mutual information \cite{Wells1996, Cizek2004}. A slightly more advanced strategy is to alternate between reconstruction and registration. For instance, this strategy was used for PET-CT \cite{Tsai2018mic, Tsai2019petct}, super-resolution MRI with CT \cite{Heinrich2012} and multi-contrast MRI \cite{Zhu2019}. However, it is not clear if this \revise{alternating procedure} converges \revise{to a limiting reconstruction / registration} and if so \revise{how the limit can be interpreted.}

Related to our proposed approach is blind deconvolution \cite{bungert2018blind,bungert2018robust} which can correct for translations between side information and measured data. However, it cannot correct for more complicated transformations.

In \cite{Fu2020} the authors propose the fusion of hyperspectral image data with a misaligned RGB image using a joint objective functional similar to \eqref{eq:joint_prob} and optimize it using PALM \cite{bolte2014PALM}. In contrast to our work, they assume that both modalities are connected via a system of linear equations which is too much of an assumption for general modalities.

\revise{Similarly, in \cite{chen2015sirf} a variational model for the fusion of misaligned multispectral and panchromatic images was proposed and solved with an alternating minimization approach.}

In \cite{chen2018indirect,lang2019template} the related problem of simultaneously reconstructing an image while registering it to a template was studied. Models to jointly reconstruct and estimate the motion between a sequence of images observed though indirect measurements were proposed in~\cite{burger2018variational,corona2019variational}. However, these methods can only deal with data from a single modality. Registering the reconstruction to a template of a different modality is not meaningful, hence we enforce registration indirectly through the regularization functional $\RegIm(u;\sinfo)$.

\revise{There are only a few neural network based approaches in the literature for reconstruction with structural priors. We would like to refer to~\cite{yang2017pannet,yuan2018multiscale} for pansharpening algorithms using deep neural nets. However, they do not consider misalignments between the panchromatic image and the multi-/ hyperspectral data.}

\section{Mathematical Model}
\label{s:model}
\subsection{Notation}
\label{ss:notation}
In this section we thoroughly define all quantities which are involved in model~\eqref{eq:joint_prob} above. The optimization takes place over images $u\in\imspace$ and deformation fields $\vf\in\vecspace$. We assume that an image $u\in\imspace$ has its pixels located at some fixed positions $x_i$ in $\R^d$ for $i=1,\ldots,n$, where $d\in\N$ denotes the image dimension (typically $d\in\{2,3\}$), which allows us to identify $\imspace\cong\R^n$ where $u_i:=u(x_i)$. Furthermore, this lets us identify a deformation field $\vf\in\vecspace$, which we denote as mapping $\vf:\R^d\to\R^d$, $x\mapsto\vf(x)$, with an array in $\R^{n\times d}$ by letting the $i$-th row be given by $\vf_i:=\vf(x_i)\in\R^d$ for $i=1,\ldots,n$. For $u\in\imspace$ and $\vf\in\vecspace$ we define their composition as 
\begin{align}\label{eq:comp_op}
u\circ\vf:=\interp(u;\vf),
\end{align}
where $\interp:\imspace\times\vecspace\to\imspace$ denotes an interpolation operator. Mathematically speaking, being an interpolation operator requires two properties: firstly, it shall hold $\interp(u;\id)=u$ for all $u\in\imspace$ where $\id\in\vecspace$ denotes the identity deformation field $\id(x)=x$. Secondly, for fixed $\vf\in\vecspace$ the map $u\mapsto \interp(u;\vf)$ shall be linear. In contrast, for fixed $u\in\imspace$ the map $\vf\mapsto \interp(u;\vf)$ is nonlinear, in general. However, we assume that $\interp$ is continuously differentiable in its second argument, which is the case for biquadratic (-cubic, -quartic, etc.) spline interpolation, for instance. Note that linear interpolation \emph{is not} differentiable and is hence not considered here. The differentiability assumption is not only of mathematical relevance but is also important from an algorithmic point of view (cf.~\cite{modersitzki2009fair}) since it assures that optimization algorithms based on derivatives are well-defined.

The forward operator in \eqref{eq:joint_prob} is a linear map $A:\imspace\to\R^m$ and $f\in\R^m$ is the measured data. In some applications, e.g. MRI, one needs a complex-valued data space. This can be incorporated in our setting by identifying the complex numbers with $\R^2$. For defining the regularization functional $\RegIm$ in~\eqref{eq:joint_prob} we will also need the notion of a gradient $\nabla u\in\vecspace$ of an image $u\in\imspace$. To this end we use forward finite differences, as detailed for instance \cite{bungert2018blind}. Throughout the paper, we let $\norm{x}$ denote the standard Euclidean norm of a vector $x$.

\revise{%
\subsection{Data fidelities}
The choice of the data fidelity $\fid$ in~\eqref{eq:joint_prob} typically depends on the statistical noise modelling of the data acquisition, cf. \cite{Scherzer2008book} and references therein. For Gaussian noise---which models the noise in MRI or hyperspectral imaging, for instance---one uses the squared Euclidean norm
\begin{align}\label{eq:L2-fid}
    \fid(u;f)=\frac{1}{2}\norm{u-f}^2.   
\end{align}
For Poisson distributed noise---which occurs during PET measurements, for instance---one utilizes the Kullback--Leibler distance 
\begin{align}\label{eq:KL-fid}
   \fid(u;f)=\int_\Omega u-f+f\log\left(\frac{f}{u}\right)\d x.
\end{align}
There are also relevant non-differentiable fidelities, as for instance, the $1$-norm $\fid(u;f)=\norm{u-f}_1$ which is used for the removal of impulse noise. However, they do not fit into our optimization framework (cf.~Section~\ref{ss:optimization}) which requires a differentiable fidelity term.}

\subsection{Structure-promoting regularization}
\label{ss:structure_regularization}
The main challenge of our model is that it combines data from different modalities. In particular, the image $u$, observed through the data $f$, and the available side information $\sinfo$ typically have completely different contrasts. This makes the methods~\cite{burger2018variational,chen2018indirect,lang2019template,corona2019variational} inapplicable. We only assume structural similarities on $u$ and $\sinfo$. While there are also other approaches~\cite{Bowsher2004}, we model this similarity in terms of shared image edges. Mathematically, this means that $u$ and $\sinfo$ ought to have parallel gradients, which can be expressed as
\begin{align}\label{eq:parallel_gradients}
\nabla u_i - \frac{\nabla\sinfo_i \nabla \sinfo_i^T}{\|\nabla\sinfo_i\|^2} \nabla u_i &= 0
\end{align}
for all $i=1,\ldots,n$. Since gradients are orthogonal to level sets, this condition is also referred to as \emph{parallel level sets}, see~\cite{Ehrhardt2014tip, Ehrhardt2015petmri, Ehrhardt2020chapter}.
Note that the left hand side in~\eqref{eq:parallel_gradients} vanishes whenever $\nabla u$ and $\nabla \sinfo$ are collinear and is equal to $\nabla u$ if they are orthogonal. To enforce this constraint, one defines the \emph{directional total variation} (cf.~\cite{Ehrhardt2016mri}) with respect to $\sinfo\in\imspace$ as
\begin{align}\label{eq:dtv}
\dtv(u;\sinfo)&= \sum_{i=1}^n \|P_i \nabla u_i\|, \\
\label{eq:projection}
P_i &=  \mathbb{1} - \xi_i \xi_i^T,\\
\label{eq:dtv_vectorfield}
\xi_i &= \gamma \frac{\nabla \sinfo_i}{\sqrt{\|\nabla\sinfo_i \|^2 + \varepsilon^2}},
\end{align}
where $\mathbb{1}$ denotes the $d\times d$ identity matrix. The parameter $\gamma\in[0,1)$ steers the influence of the side information from no ($\gamma=0$) to high ($\gamma\approx 1$), where $\gamma=1$ should be avoided for theoretical reasons, see~\cite{Hintermuller2018,bungert2018blind}. For $\gamma=0$ one observes that $\dtv$ reduces to the standard total variation $\tv$. Furthermore, $\varepsilon>0$ is a small parameter which assures that~\eqref{eq:dtv_vectorfield} is well-defined even if~$\nabla\sinfo_i=0$. Note that if $\gamma=1$ and $\varepsilon=0$, then~\eqref{eq:projection} is the projection onto the orthogonal complement of $\mathrm{span}(\nabla\sinfo_i)$ and in this case~\eqref{eq:parallel_gradients} holds for all $i = 1, \ldots, n$ if and only if $\dtv(u;\sinfo) = 0$. \revise{Conversely, choosing $\varepsilon>0$ and $\gamma<1$ implies that $\nabla u_i=0$ is energetically cheaper for $\dtv$ than $\nabla u_i=\nabla v_i$. This ensures that no artificial edges of $v$ are introduced.}

For many modalities the physical assumption that images are nonnegative makes sense. In these cases, we enforce this via the characteristic function 
\begin{align}
\iota_+(u)=
\begin{cases}
0,\quad&\text{if }u_i\geq 0\quad\forall i=1,\ldots,n,\\
\infty,\quad&\text{else},
\end{cases}
\end{align}
and set the regularization function $\RegIm$ in~\eqref{eq:joint_prob} as
\begin{align}\label{eq:defi_reg_im}
\RegIm(u; v):=\alpha\dtv(u;\sinfo)+\iota_+(u).
\end{align}
Here, $\alpha>0$ denotes a regularization parameter which steers the amount of regularization used.

We would like to stress that the overarching framework of this contribution is not limited to this particular regularizer and can be used in conjunction with other models, too, for example \cite{Leahy1991, Bowsher2004, Ballester2006, Bilgic2011, Moeller2012, Huang2014multicontrastmri, Mehranian2017petplusmri, Rasch2018, Song2020mri, Ehrhardt2020chapter}.

\subsection{Parametric deformations}
\label{ss:parametric}

In practice, frequently the kind of deformation is approximately known. For instance, a CT and an MRI image of the brain of the same patient can be assumed to be connected by an affine transform of the form
\begin{equation}\label{eq:affine_def}
    \vf(x)=
    Mx+b,\quad x\in\R^d.
\end{equation}
where $M\in\R^{d\times d}$ and $b\in\R^d$. Special cases include \emph{rigid motions} in the plane with rotation matrix
\begin{align}
    M=R_\theta:=
    \begin{pmatrix}
    \cos\theta & -\sin\theta \\
    \sin\theta & \cos\theta
    \end{pmatrix}, \quad \theta\in(0,2\pi] \label{eq:rotation}
\end{align}
and \emph{shears} with the shear matrix
\begin{align}\label{eq:shear}
    M = S_a := 
    \begin{pmatrix}
    1 & a\\ 
    0 & 1
    \end{pmatrix}  , \quad a \in \R.  
\end{align}
However, one can also think of higher-order parametrizations of the form $\vf(x)=H(x,x)+Mx+b$ for $x\in\R^d$ where $H:\R^d\times\R^d\to\R^d$ denotes a bilinear form parametrized by a tensor with $d^3$ degrees of freedom. For more examples of parametric deformations we refer to \cite{modersitzki2009fair}. For a review on different methods in image registration we refer to \cite{modersitzki2009fair,oliveira2014medical}. An advantage of using parametrized deformations is that the number of variables in the optimization problem~\eqref{eq:joint_prob} is dramatically reduced since not a whole deformation field with $n\times d$ entries has to be estimated but only the parameters of the parametrization. Remember that $n$ denotes the number of pixel which can be very large, and $d$ denotes the dimension of the physical space which is typically $2$ or $3$. In the case of affine transformations~\eqref{eq:affine_def} the number of parameters is $d^2+d=6$ for $d=2$ and for rigid motion in the plane it is only $3$. Hence, in these cases the number of parameters is significantly smaller than $n\times d$. Moreover, parametric deformations do not require sophisticated regularization, as used for instance in \cite{burger2018variational,corona2019variational}.

We incorporate the parametrization through a map $P$ from the parameter space $\R^p$ to the space of deformation fields $\vecspace$. In general, $P$ can be nonlinear---as it is the case for rigid motion, for instance---but for affine deformations~\eqref{eq:affine_def} it is linear and given by $P:\R^6\to\vecspace$, defined as
\revise{%
\begin{align}\label{eq:param_affine}
    P(\pvf)=
    \left[
    x\mapsto
    \begin{pmatrix}
    1+\pvf_1 & \pvf_2 \\
    \pvf_3 & 1+\pvf_4
    \end{pmatrix}
    x+
    \begin{pmatrix}
    \pvf_5\\
    \pvf_6
    \end{pmatrix}
    \right]
\end{align}
in two spatial dimensions. Note that the parameters $\pvf$ model the \emph{deviation from the identity} such that $P(0)=\id$.} For a more compact notation we suppress the dependency on the parametrization $P$ and write
\begin{align}\label{eq:comp_with_parametrization}
    u_\pvf:=u\circ P(\pvf)=\interp(u;P(\pvf)),
\end{align}
where we used the composition operator~\eqref{eq:comp_op}. Furthermore, incorporating parametric deformations into~\eqref{eq:joint_prob} we obtain
\begin{align}\label{eq:actual_model}
    \min_{u\in\imspace, \pvf\in\R^p}
    \fid(\fwd u_\pvf ; f)+\RegIm(u;\sinfo)+\RegVf(\pvf).
\end{align}

\section{Algorithmic Framework}
\label{s:algortihm}
\subsection{Optimization}\label{ss:optimization}
\begin{algorithm}
\caption{Proximal alternating linearized minimization (PALM) to solve~\eqref{eq:generic_prob}.}\label{alg:palm}
\begin{algorithmic}[1]
\renewcommand{\algorithmicrequire}{\textbf{Input:}}
\renewcommand{\algorithmicensure}{\textbf{Output:}}
\REQUIRE $u,\,\pvf,\,\sigma>0,\,\tau>0,\,K \in \N$ \\
\ENSURE $u, \pvf$
\FOR {$k = 1, \ldots, K$}
\STATE $u=\prox_{\sigma\RegIm}\left(u-\sigma\partial_u\coupling(u,\pvf)\right)$ \\
\STATE $\pvf=\prox_{\tau\RegVf}\left(\pvf-\tau\partial_\pvf\coupling(u,\pvf)\right)$ \\
\ENDFOR
\end{algorithmic} 
\end{algorithm}

In this section, we detail the optimization scheme that we use to solve problem~\eqref{eq:actual_model} numerically. Before we formulate the algorithm, we cast problem~\eqref{eq:actual_model} into the form 
\begin{align}\label{eq:generic_prob}
\min_{\substack{u, \pvf}}\coupling(u,\pvf)+\RegIm(u)+\RegVf(\pvf),
\end{align}
where $\coupling(u,\pvf)=\fid(\fwd u_\pvf ; f)$ denotes the data fidelity term and we abbreviated $\RegIm(u)=\RegIm(u;\sinfo)$. Note that---due to the smoothness assumption on the interpolation operator in~\eqref{eq:comp_op}---the function $\coupling$ is smooth in both variables \revise{if the fidelity $\fid$ is}. Furthermore, the objective function in~\eqref{eq:generic_prob} is non-convex in the joint variable $(u,\pvf)$ and also non-convex in $\pvf$. However, it is convex in the variable $u$. Still, due to the overall non-convexity one can in general only expect to find critical points of~\eqref{eq:generic_prob} using gradient-based algorithms and the outcome depends on the initialization of the numerical optimization algorithm. Another difficulty arises from the non-smoothness of the regularization functionals in~\eqref{eq:generic_prob} which excludes many gradient-based algorithms. Here we employ the proximal alternating linearized minimization (PALM) algorithm \cite{bolte2014PALM} which amounts to alternating forward-backward splitting, see Algorithm~\ref{alg:palm}.

The proximal operator \cite{Bauschke2011, Parikh2014, Chambolle2016actanumerica} of a proper function $\mathcal J:\R^k\to\R\cup\{\infty\}$ is defined as
\begin{align}
\prox_{\mathcal J}(x):=\argmin_{y\in\R^k}\left(\frac{1}{2}\norm{x-y}^2+\mathcal J(y)\right).
\end{align}
The proximal operator for $\dtv$ with nonnegativity constraint \eqref{eq:defi_reg_im} can be computed using the fast gradient projection algorithm (also known as FISTA) \cite{Beck2009}, see \cite{Ehrhardt2016mri} for  details.

Let us now study the gradients of the data fidelity used in Algorithm~\ref{alg:palm}. With the chain rule and~\eqref{eq:comp_with_parametrization}, these are given by
\begin{align}%
\label{eq:grad_coupling_u}
\partial_u\coupling(u,\pvf)&=\partial_u \interp(u;P(\pvf))^*\fwd^* \partial\fid(\fwd u_\pvf; f) \\
\label{eq:grad_coupling_vf}
\partial_\pvf\coupling(u,\pvf)&=\partial P(\pvf)^*\partial_\vf \interp(u;P(\pvf))^*\fwd^*  \partial\fid(\fwd u_\pvf; f) 
\end{align}
where the asterisk denotes the adjoint / transpose of a linear map. Note that since the interpolation operator $\interp$ is linear in the first argument, it holds that $\partial_u\interp=\interp$ and its adjoint---which involves inverting the deformation---can be explicitly calculated for affine deformations. In our implementation we use the \texttt{scipy} \cite{virtanen2020scipy} routine \texttt{RectBivariateSpline} which allows to evaluate bivariate splines of order two or higher together with their derivatives~$\partial_\vf\interp$. Furthermore, the linear operator $\partial P(\pvf)^*$ can be explicitly calculated in dependency on the type of parametrization used. For affine deformations, for instance, map $P$ in~\eqref{eq:param_affine} is linear; hence, it holds $\partial P=P$ and the adjoint map $P^*:\vecspace\to\R^6$ can be computed easily.

Finally, we address the choice of step sizes $\sigma$ and $\tau$ appearing in Algorithm~\ref{alg:palm}. From a theoretical point of view it suffices to choose them smaller than the reciprocal of the Lipschitz constants of the gradient maps in~\eqref{eq:grad_coupling_u}, \eqref{eq:grad_coupling_vf} to obtain convergence of the algorithm. However, in practice those Lipschitz constants are hard to compute analytically and typically one can only find pessimistic upper bounds for them. Since this slows down convergence, we employ a backtracking scheme instead, see \cite{bungert2018blind} for details.

\subsection{Initialization strategy} \label{ss:initialization}
Since image registration alone is already a severely non-convex problem, one can expect that the degree of non-convexity of our joint model~\eqref{eq:joint_prob} is even higher. In particular, the outcome of Algorithm~\ref{alg:palm} will depend heavily on its initialization, in general. Hence, in order to be able to detect large deformations between the side information $v$ and the image $u$ such that $Au \approx f$, we employ a scale-space strategy commonly used in image registration~\cite{modersitzki2009fair}. To overcome large deformations we reduce the resolution of the problem such that these correspond to a few pixels only. We also employ an over-smoothing strategy based on the regularization parameter $\ParamIm$ in~\eqref{eq:defi_reg_im}. It is well-known for a large class of regularization functionals of the form $\RegIm=\ParamIm\mathcal{J}$, that solutions of the associated variational regularization method converge to an element in the null-space of $\mathcal{J}$ as $\ParamIm\to\infty$ (see \cite{burger2013guide} for total variation and \cite{bungert2019solution} for the general case). For the case of directional total variation, i.e. $\mathcal{J}=\dtv$, this means that for high regularization parameters $\RegIm$ the reconstructed images become less oscillatory and better align with the side information. However, to avoid a strong loss of contrast in the reconstructions~\cite{burger2013guide}, one would like to choose the regularization parameter as small as possible. In this work we propose a combination of resolution-based and over-smoothing strategy to solve~\eqref{eq:joint_prob} which is detailed in Algorithm~\ref{alg:scale}. In a nutshell, the algorithm first reconstructs low-resolution images with high regularization parameters first and then successively decreases the regularization parameter while reconstructing better resolved images.

\begin{algorithm}
\caption{Scale-space strategy to solve~\eqref{eq:generic_prob}.}\label{alg:scale}
\begin{algorithmic}[1]
\renewcommand{\algorithmicrequire}{\textbf{Input:}}
\renewcommand{\algorithmicensure}{\textbf{Output:}}
\REQUIRE  $u,\,\pvf$, resolutions $n_1 < \ldots < n_M = n$, \\ regularization parameters $\alpha_1 > \ldots > \alpha_M$ in~\eqref{eq:defi_reg_im} \\
\ENSURE $u, \pvf$
\FOR {$i = 1, \ldots, M$}
\STATE down-sample $\sinfo$ to resolution $n_i$ \\
\STATE up-sample $u$ to resolution $n_i$ \\
\STATE compute $(u, \pvf) = \mathrm{PALM}(u, \pvf)$ \\
\ENDFOR
\end{algorithmic} 
\end{algorithm}

\section{Numerical results}
\label{s:numerics}
In all numerical experiments we will use affine parametric deformations (cf.~Section~\ref{ss:parametric}), since they model many realistic deformations and are computationally efficient due to the low number of free parameters. Furthermore, we do not regularize the affine deformation fields, meaning that we choose $\RegVf(\pvf)=0$ for all $\pvf\in\R^p$ in \eqref{eq:actual_model}. Correspondingly, the second proximal operator in Algorithm~\ref{alg:palm} reduces to the identity operator. The details of the scale-space Algorithm~\ref{alg:scale} are specified below but at this point we already remark that we use $K=500$ iterations of the PALM algorithm in each loop of Algorithm~\ref{alg:scale}. The model parameters in~\eqref{eq:dtv_vectorfield} were chosen similar to~\cite{Ehrhardt2016mri, bungert2018blind} as $\gamma=0.9995$, $\varepsilon=0.01 \cdot \max_{i=1}^n\norm{\nabla v_i}$. The numerical experiments have been carried out in Python using ODL \cite{Adler2017odl}. The line integrals in the PET experiment were computed using the Python module \texttt{scikit-image}~\cite{van2014scikit} and for interpolation we use biquadratic splines from the \texttt{scipy}~\cite{virtanen2020scipy} routine \texttt{RectBivariateSpline}. {The source code which reproduces all experiments can be found on \url{https://github.com/mehrhardt/robust_guided_reconstruction}.}

\revise{%
The error metrics we use in this section are structural similarity (SSIM) \cite{wang2004image} for the comparison of reconstructed and ground truth images, and the relative difference (RD), given by
\begin{align}\label{eq:relative_diff}
    \mathrm{RD} = \frac{\norm{\pvf-\pvf_\mathrm{gt}}}{\norm{\pvf_\mathrm{gt}}}\times 100\, [\%],
\end{align}
for the comparison of reconstructed motion parameters $\pvf$ with ground truth parameters $\pvf_\mathrm{gt}$. Whereas more advanced evaluation metrics (cf. e.g.~\cite{seitzer2018adversarial} for semantic evaluation of neural networked based MR reconstruction) would be desirable, they are beyond the scope of this paper and the combination of SSIM, RD, and visual assessment appears to be sufficient for our applications.}

In our numerical experiments we compare our method with a three-step method which computes an initial reconstruction, registers the side information, and subsequently reconstructs again (cf.~the references in Section~\ref{ss:related_work}). 
Based on the structural side information $v$ and data $f$ it computes \revise{%
\begin{subequations}\label{eq:alternating}
\begin{align}
    \tilde{u} &\in\argmin_{u\in\imspace}\fid(\fwd u;f)+\ParamIm\tv(u),\\
    \label{eq:registration_step}
    \pvf^* &\in \argmax_{\pvf\in\R^p}\mathcal{MI}(v,\tilde{u}_\pvf), \\
    u^* &\in\argmin_{u\in\imspace}\fid(\fwd u_{\pvf^*};f)+\ParamIm\dtv(u;v).
\end{align}
\end{subequations}}
The maximum is taken over all parametrized affine deformation fields $\pvf\in\R^p$.
Furthermore, $\mathcal{MI}(\cdot,\cdot)$ denotes the mutual information~\cite{pluim2003mutual} of images which is a popular distance measure for the registration of multi-modal images.
\revise{%
The affine registration step \eqref{eq:registration_step} was performed using the MATLAB$^\copyright$ routine \texttt{imregtform}.
}

\subsection{Undersampled multi-contrast MRI reconstruction} \label{ss:mri}
\input{figures/fig_3}
In our first experiment we perform multi-contrast MRI reconstruction from undersampled data. Since all contrasts are likely to share a structural information, less data is needed if the expected redundancy is exploited. Here we assume that a T2-weighted has been acquired using "full data" and reconstructed without artefacts. We then use this image to reconstruct a T1-weighted image using only a fraction of the data. \revise{In this experiment we simulate the k-space data based on reconstructions of fully sampled clinical data.}

The forward operator is a discrete Fourier transform defined on complex-valued images of size $256\times 256$, followed by a sampling operator corresponding to 15 equidistant spokes \revise{and a low-pass sampling of the $10\times 10$ center of the $k$-space}. This yields a sampling of approximately $3\%$ of the $k$-space. \revise{This sampling is easy to approximate on any clinical MRI scanner. The noise modelling here is Gaussian which is why we use the squared Euclidean norm \eqref{eq:L2-fid} as data fidelity.} The image domain is $[-1,1]^2$ which corresponds to a pixel width of approximately $0.0078$. The deformation field in this simulation is a rigid transform followed by zooming, given by
\begin{align*}
    \vf_\texttt{zoom}(x)=
    \alpha R_\theta x + b
\end{align*}
where $R_\theta$ is a rotation matrix~\eqref{eq:rotation} with angle $\theta=0.1$, $b=(-0.02,-0.08)^T$ is a translation vector, and $\alpha=0.85$ is a zoom factor. The latter means that the size of the side information image is only $85\%$ of the size of the image which underlies the $k$-space data. Since nonnegativity is not a meaningful assumption for complex-valued images, we drop this constraint in \eqref{eq:defi_reg_im} for this test case. The resolutions and regularization parameters in Algorithm~\ref{alg:scale} were chosen as $n_k=(32^2,62^2,128^2,256^2)$ and $\alpha_k=5\cdot 10^{-3}\cdot(5^3,5^2,5,1)$. For the $\tv$ reconstruction we used $\alpha_k=5\cdot 10^{-4}\cdot(5^3,5^2,5,1)$. \revise{Here, and also for the following experiments, the resolutions were determined by successively dividing the resolution of the side information by a factor of two. The regularization parameters at the target resolution were chosen experimentally in order to maximize SSIM. Typically, the regularization parameter of $\tv$ has to be chosen one or two orders of magnitude smaller than the one for $\dtv$.}

The first row of Figure~\ref{fig:MRI_comparison} shows the sampling pattern and the deformed image which was used to generate the data. Furthermore, it shows the side information and the ground truth image which we would like to reconstruct. 
 
In the second row of Figure~\ref{fig:MRI_comparison} we show the zero-filled reconstruction (pseudo-inverse), a standard $\tv$ reconstruction, the result of the three-step method~\eqref{eq:alternating}, and our proposed method. Obviously the pseudo-inverse and $\tv$ method yield unsatisfactory results since the do not utilize the side information.  Furthermore, they visualize the strong degree of ill-posedness of the problem where the sampled Fourier data alone cannot be used for a meaningful reconstruction. 
In contrast, both the three-step method~\eqref{eq:alternating} and our proposed method successfully correct the deformation between side information and data and yield good reconstructions despite the high degree of undersampling in the data. 
Notably, the proposed method yields better reconstructions of fine structures as can be seen in the zooms of Figure~\ref{fig:MRI_comparison}.

\revise{%
The previous observations are also supported by quantitative metrics. The SSIM values between the pseudo-inverse and $\tv$ reconstruction and the deformed ground truth image which underlies the data are very low which is due to the strong undersampling. In contrast, the three-step method and the proposed method have high similarity values with the ground truth image, with slightly better numbers of the proposed method. Also the relative difference between the estimated affine deformation parameters and the ground truth parameters is lower for the proposed method than for the three-step method which underlines the superiority of our joint approach over the three-step method~\eqref{eq:alternating} in this application. We suspect that worse estimation of the deformation field of the three-step method is due to the fact that it must utilize the blurry and blocky $\tv$ reconstruction for registration whereas the proposed method iteratively improves registration and reconstruction.}
\input{figures/fig_4}
\input{figures/fig_5}
We visualize our algorithm in Figure~\ref{fig:MRI_scale_space} where we plot the reconstructions and deformations fields of the scale-space Algorithm~\ref{alg:scale}, respectively. \revise{Since our reconstructed motion parameters correspond to the deviation from the identity (see~\eqref{eq:param_affine}), we let the yellow arrows indicate the field $\vf_\mathrm{zoom}-\id$ and the red arrows the corresponding expression for the reconstructed fields.}
\revise{%
Already at the second resolution stage $60\times 60$ the estimated deformation field is visually not distinguishable from the ground truth field which is also supported by the decreasing relative differences between the estimated and ground truth motion parameters. 
}
\subsection{PET-MR reconstruction}
\label{ss:pet}
\input{figures/fig_6}
\revise{In this experiment we consider PET-MR, where we aim to reconstruct a tracer distribution using a fully sampled T1-weighted MR image of size $144\times 144$ as side information.}
The forward operator is modelled by a parallel beam X-ray transform with 200 angles equispaced in $(0,\pi]$ and 192 bins. 
The sinogram data were simulated using a ground truth image deformed with respect to the side information through the rigid deformation
\begin{align}
    \vf_\rigid(x)=R_\theta x+b, \label{eq:rigid}
\end{align}
where $R_\theta$ is a rotation matrix~\eqref{eq:rotation} with angle $\theta=0.1\approx 5.7^\circ$ and $b=(0.02,0.08)^T$ is a translation vector. 

\revise{In this experiment we simulate data based on a $\dtv$-regularized reconstruction of clinical data, see \cite{Ehrhardt2019pmb}. The data is an instance of a Poisson random variable with parameter $Ax + r$, where the  background $r$ is chosen as constant 7 and the forward operator is scaled to about $1.3\cdot 10^6$ expected counts in the data. Correspondingly, the data fidelity used is the Kullback--Leibler distance~\eqref{eq:KL-fid}.

Again, images are in $[-1,1]^2$ and hence the pixel width of the side information is $0.013\overline{8}$. The resolutions and regularization parameters in Algorithm~\ref{alg:scale} were chosen as $n_k=(9^2,18^2,36^2,72^2,144^2)$ and $\ParamIm_k=4\cdot10^{-1}\cdot(10^4,10^3,10^2,10,1)$. For the $\tv$ experiment we used $\ParamIm_k=4\cdot10^{-2}\cdot(10^4,10^3,10^2,10,1)$.}

The sinogram data and the deformed image which was used to generate the data are shown in the top row of Figure~\ref{fig:PET_comparison}. Furthermore, we show the side information and the ground truth image. In the second row of Figure~\ref{fig:PET_comparison} we show four different reconstructions: the first one obtained through  filtered back-projection, the second one utilizing $\tv$ regularization, the third one using the three-step method~\eqref{eq:alternating}, and the fourth one being the proposed method. The first two methods, which do not use the side information or correct any motion, exhibit poor image quality due to strong noise in the sinogram.
\revise{%
On the other hand, both the three-step and the proposed method correct the deformation and the reconstructions are in very good agreement with the ground truth image.

Quantitatively, the SSIM values for filtered back-projection and $\tv$ are relatively low whereas they are comparably high for both the three-step and the proposed method, with slightly better values for the proposed method. The same is also true for the relative errors of the computed deformation fields.}

The scale-space generated by Algorithm~\ref{alg:scale} is again visualized in Figure~\ref{fig:PET_scale_space} with observations similar to Section~\ref{ss:mri}.

We also test the limits of the three-step method and our method by performing PET reconstructions where the measured sinogram data and the side information at hand are offset by the rigid deformation $\phi_\rigid$ from \eqref{eq:rigid} with \revise{$\theta\in\{0.2,0.4,\dots,1.2\}$.}
\revise{%
The first row of Figure~\ref{fig:rotations} shows the deformed ground truth images which were used to generate the PET data. In the second row we show the reconstructions generated by the three-step method~\eqref{eq:alternating} together with SSIM and RD values for the images and deformation parameters, respectively. 
The remaining five rows of Figure~\ref{fig:rotations} show the results of our proposed method for different sizes $M\in\{1,2,3,4,5\}$ of the scale space in Algorithm~\ref{alg:scale}. The method behaves very reliably in the sense that a larger scale space implies enables the method to estimate larger deformations correctly. In particular, for scale space of size $3$ or larger our method breaks down much later than the three-step method~\eqref{eq:alternating} which internally uses state-of-the-art multimodal registration algorithms. This suggests a high degree of robustness of our method and also shows that the scale space algorithm~\ref{alg:scale} is inevitable for joint reconstruction~/ registration problems.
Also the error metrics SSIM and RD are better for the proposed method than for the three-step method, as it was the case in the previous experiments.}
\input{figures/fig_7}

\subsection{Hyperspectral super-resolution}
\label{ss:superresolution}
In this section we fuse a $100\times 100$ spectral data channel with a highly resolved panchromatic side information of size $400\times 400$. \revise{We assume Gaussian noise and hence utilize the Euclidean fidelity~\eqref{eq:L2-fid}.}

We consider three test cases \rigid{}, \shear{}, and \nonlinear{} where we artificially introduce different deformations between data and side information. For \rigid{} we utilize the deformation $\vf_\rigid$ in~\eqref{eq:rigid}
with $\theta=0.1\approx 5.7^\circ$ and $b=(0.06,-0.04)^T$. The pixels of the images lie in $[-1,1]^2$ and have width $0.02$ (data image) and $0.005$ (side information). Hence, we try to correct an effective off-set of approximately 12 pixels. The other deformations are given by $\vf_\shear(x)=S_a x+b$---where $S_a$ is the shear matrix~\eqref{eq:shear} with $a=0.08$---and
\begin{align*}
    \vf_\nonlinear(x)&=\vf_\rigid(x)+0.05
    \begin{pmatrix}
    x_2^2\\
    -x_1^3
    \end{pmatrix}.
\end{align*}
We notice that $\vf_\nonlinear$ is a nonlinear perturbation of $\vf_\rigid$ and, in particular, is \emph{not affine}. In the scale-space Algorithm~\ref{alg:scale} we chose five different resolutions $n_k=(25^2,50^2,100^2,200^2,400^2)$ and corresponding regularization parameters $\ParamIm_k=10^{-3}\cdot (10^4,10^3,10^2,10,1)$. These parameters were used for all reconstructions involving $\dtv$ and for all test cases. In the experiment with standard total variation $\tv$ as the regularizer we used the parameters $\ParamIm_k= 10^{-5}\cdot (10^4,10^3,10^2,10,1)$ instead.

The first row of Figure~\ref{fig:RS_comparison} shows the data \rigid{} together with the side information  misaligned through a rigid transform. Below we show four different reconstructions, including a target reconstruction which was computed using aligned data and serves as substitute for a ground truth solution for this real data set. The second one is a standard $\tv$ reconstruction, which does not utilize the side information at all and hence yields the expected poor and blurry super-resolution result.
Instead of the three-step method---which we compared with our method in Sections~\ref{ss:mri} and \ref{ss:pet}---the third reconstruction is now a $\dtv$ reconstruction which uses misaligned data and side information and does not correct the deformation. 
This result corresponds to a naive super-resolution which does not take into account the misalignment between the two modalities and hence produces an erroneous image with many artefacts.
The last reconstruction is obtained using our proposed method which corrects for the misalignment between data and side information and almost perfectly agrees with the target reconstruction which is also reflected by its SSIM value.
\input{figures/fig_8}
In Figure~\ref{fig:RS_recons} we apply our method also to the other data sets \shear{} and \nonlinear{}. Also here our algorithm was able to correct the deformation between data and side information sufficiently to achieve reconstructions which are very similar to the target reconstruction, as also shown by SSIM.
Note that even though the deformation field in the data set \nonlinear{} is not affine, the reconstruction is almost as good as for the affine data sets \rigid{} and \shear{} and the SSIM to the target reconstruction is only slightly lower than for \rigid{} or \shear{}. Since the ground truth deformation field is non-affine for this data set, it is not meaningful to provide relative distances of the estimated parameters here.
\input{figures/fig_9}
\section{Discussion}\label{s:discussion}
\revise{%
In this work we have proposed a variational framework for jointly reconstructing an image from indirect measurements while registering it to a structural side information from a potentially different modality. We have then specialized our model to the case of parametric deformations. In numerical experiments on multi-contrast MR, PET-MR, and hyperspectral super-resolution we have compared our method with standard reconstruction methods (pseudo inversion, total variation regularization) and with a three-step method which first reconstructs an auxiliary image, then registers the side information, and finally reconstructs again. 

While the standard methods, which do not use any structural prior, do not produce satisfactory results for the type of inverse problems we are considering, both the three-step method and the proposed method yield comparable results and their pros and cons are discussed in more detail in the following. 

The clear benefit of the three-step method is that it is straightforward to implement since it can be constructed by combining any two (black-box) reconstruction and multi-modal registration algorithms. Furthermore, by using state-of-the-art solvers of the sub-problems one can create a very efficient method. However, the drawback of this strategy becomes evident in our numerical experiments on MR and PET-MR where---due to the ill-posedness of the problems---the initial reconstruction does not suffice to estimate an accurate deformation field. Correspondingly, the final reconstructions of the three-step method is not as accurate as our proposed approach. \LB{Furthermore, as seen in Section~\ref{ss:pet} the three-step methods seems to be less robust to large deformations since it utilizes a possibly poor initial reconstruction which makes the registration hard.}

Thanks to its joint approach, which \emph{simultaneously} estimates the deformation field and the reconstruction, our proposed method is able to estimate deformations more accurately which reflected in better image quality \LB{and higher robustness to larger deformations}. Another benefit of our method is that, despite its non-convexity, it can be solved with optimization algorithms which come with convergence guarantees. 

Incorporating more general deformation models than the affine one considered here can be achieved by suitably choosing the regularization functionals $\RegIm$ and $\RegVf$ in \eqref{eq:joint_prob} or even choosing a joint regularization. For diffeomorphic deformations this can be done similarly to~\cite{corona2019variational} and for optical flow approaches see~\cite{burger2018variational}. In any case, judging from what the authors report in these works, the scale space algorithm~\ref{alg:scale} is necessary also for more complicated types of deformations.

Finally, it is also worth discussing the limitations of the assumption that side information and ground truth solution have structural similarity in terms of parallel gradients (cf.~\eqref{eq:parallel_gradients}). For the scenario of an aligned side information it was demonstrated in~\cite{Ehrhardt2015petmri,ehrhardt2016pet,rasch2017joint} that features which are only present in PET data but not in the structural MR prior can be recovered in the reconstruction. Similarly, excess edges in the MR image were shown not to propagate into the reconstruction. A comprehensive overview of different MR-informed PET reconstruction methods is given in~\cite{bland2019intercomparison}. While these observations were made for registered modalities, one should mention that in our case the registration (enforced through the regularization functional $\RegIm(u;v)$) heavily depends on the side information. In particular, a successful registration can only be obtained if most of the edges in the data are also present in the side information. In the extreme case where the side information is a constant image and no registration can be performed, this condition is of course violated.}

\section{Conclusions}\label{s:conclusion}
Multi-modality imaging is becoming ever more important in many disciplines from medicine to material sciences.  Mathematical approaches to combine data from multiple modalities exist but are prone to imperfect misalignment. In this work we proposed a generic framework to reconstruct an image from indirect measurements while registering it to a structural side information from a potentially different modality. Numerical experiments for a specific regularizer (directional total variation) and a variety of applications including hyperspectral imaging in remote sensing, PET-MR and multi-contrast MRI underpinned the aptness of the proposed approach to correct misalignments between data and structural side information which cause existing algorithms to fail. Missing robustness to misalignment has been the biggest hurdle for integrating structure promoting regularizers into routine use, for instance in the clinic. Thus, the proposed framework paves the way for their translation into various applications.

\bibliographystyle{IEEEtran}
\bibliography{bibliography}

\begin{IEEEbiography}[{\includegraphics[width=1in,height=1.25in,clip,keepaspectratio]{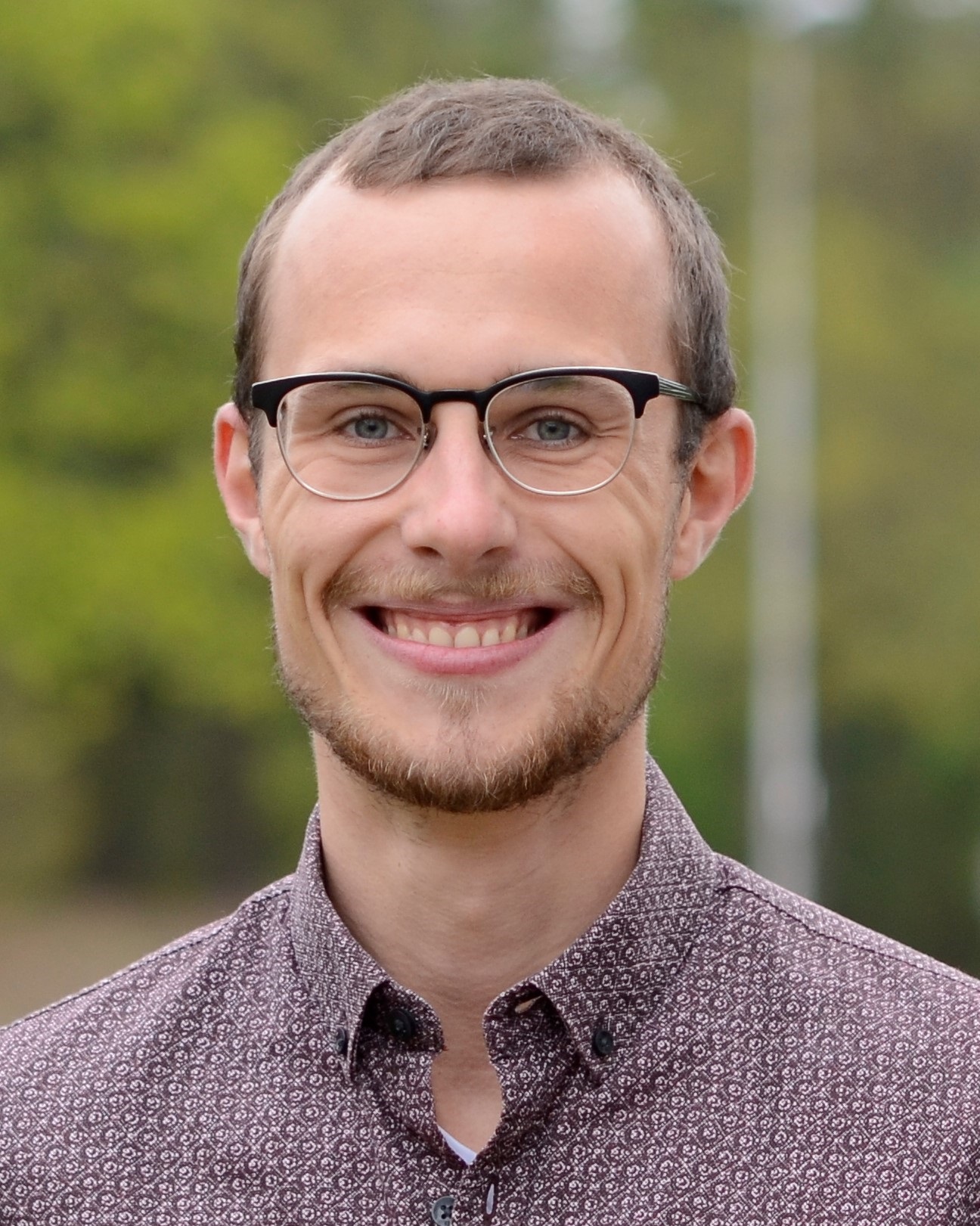}}]{Leon Bungert} obtained a Master's degree in mathematics at the University of Erlangen, Germany, in 2017. In 2020 he received his PhD \emph{summa cum laude} at the same institution. His research interests include nonlinear eigenvalue problems, partial differential equations (on graphs), inverse problems, and image reconstruction.
\end{IEEEbiography}

\begin{IEEEbiography}[{\includegraphics[width=1in,height=1.25in,clip,keepaspectratio]{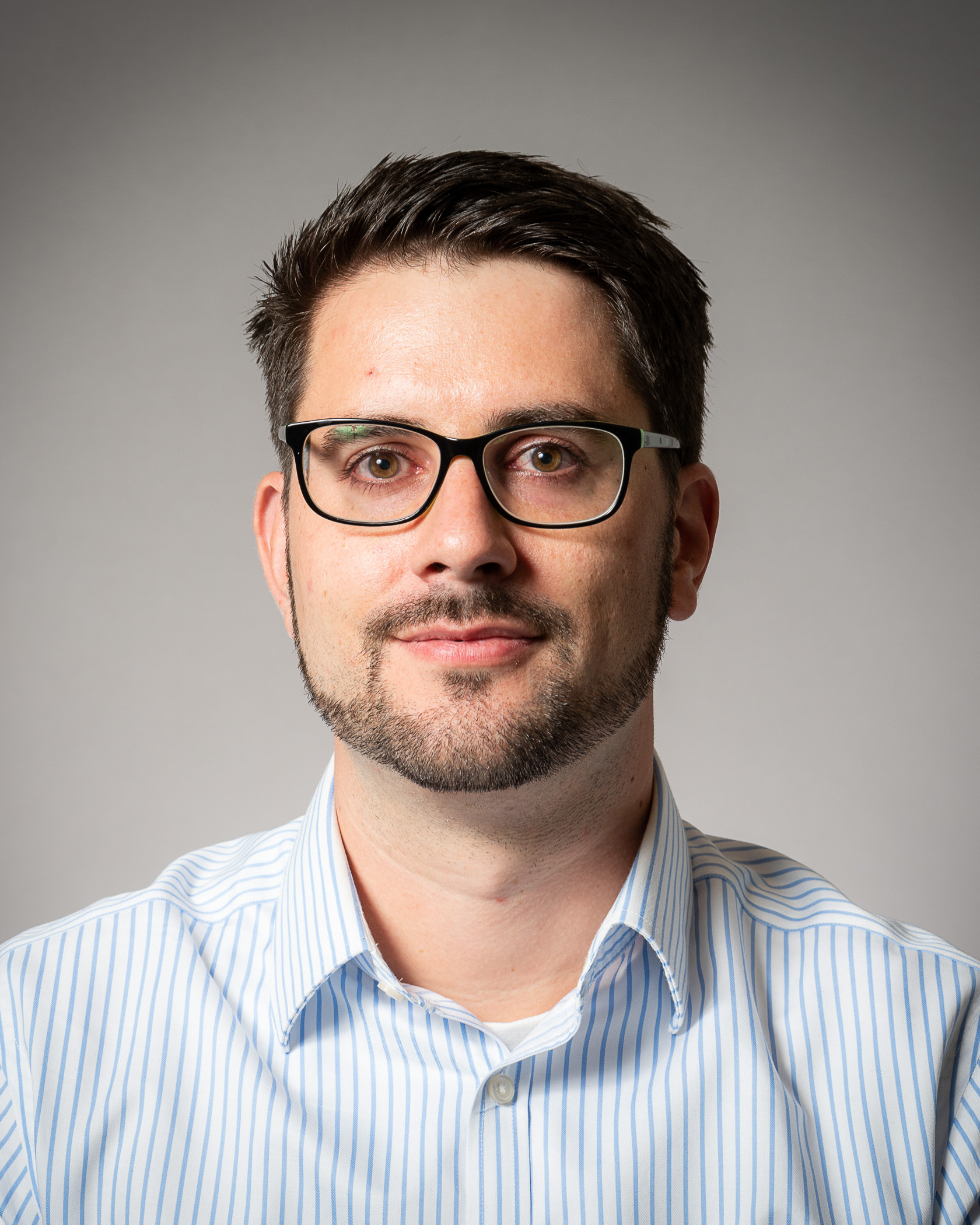}}]{Matthias J. Ehrhardt} received a Diploma with honours in industrial mathematics from the University of Bremen, Germany, in 2011, and his PhD in medical imaging from University College London, United Kingdom, in 2015. He was a postdoc in the Cambridge Image Analysis group at the Department for Applied Mathematics and Theoretical Physics at the University of Cambridge, United Kingdom, from 2016 to 2018. He is currently a Prize Fellow at the Institute for Mathematical Innovation, University of Bath, United Kingdom. His research interests include optimization, inverse problems, computational imaging and machine learning.
\end{IEEEbiography}
 
\EOD

\end{document}

%% file: figures/fig_1-2.tex
\begin{figure}%
\centering%
\def\PicHeight{1.95cm}%
\PlotFigs{\includegraphics[height=\PicHeight]{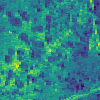}\includegraphics[height=\PicHeight]{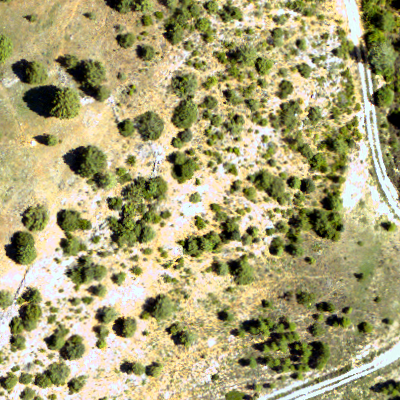}}{Hyperspectral imaging}%
\hfill%
\PlotFigs{%
\includegraphics[clip, trim=5px 0px 0px 20px, height=\PicHeight]{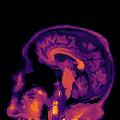}\hspace*{-1pt}%
\includegraphics[clip, trim=5px 0px 0px 20px, height=\PicHeight]{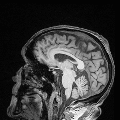}%
}{PET-MR\phantom{g}}%
\caption{In many applications two images of different contrast and resolution are acquired. Images courtesy of D. Coomes, P. Markiewicz and J. Schott.}\label{fig:motivation}%
\vspace*{5mm}%
\def\PicWidth{6cm}%
\centering%
\begin{tikzpicture}[inner sep=0pt]
\begin{scope}
\draw (-0.02, -0.02) node [anchor=south west] {\includegraphics[height=2.8cm]{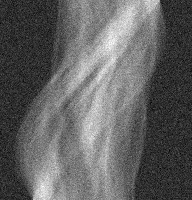}};%
\end{scope}
\node[align=left,anchor=north west] at (0.01,-0.1) {\footnotesize \color{black} PET sinogram data\\[-3pt]
\phantom{lg}};
\end{tikzpicture} \hfill%
\begin{tikzpicture}[inner sep=0pt]
\begin{scope}
\clip (0,0) rectangle (2.8cm, 2.8cm);
\draw (-3cm, -3cm) node [anchor=south west] {\includegraphics[height=8cm]{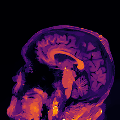}};%
\end{scope}
\node[align=left,anchor=north west] at (0.01,-0.1) {\footnotesize \color{black} reconstruction with\phantom{Pg}\\[-3pt]
\footnotesize aligned side info};
\end{tikzpicture} \hfill%
\begin{tikzpicture}[inner sep=0pt]
\begin{scope}
\clip (0,0) rectangle (2.8cm, 2.8cm);
\draw (-3cm, -3cm) node [anchor=south west] {\includegraphics[angle=5, width=8cm]{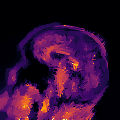}};%
\end{scope}
\node[align=left,anchor=north west] at (0.01,-0.1) {\footnotesize \color{black} reconstruction with\phantom{Pg}\\[-3pt]
\footnotesize misaligned side info};
\end{tikzpicture}%
\caption{%
\revise{While reconstruction with structural side information (center) can yield very good reconstructions even for noisy data (left), the reconstruction is completely distorted when the side information is misaligned (right).}}\label{fig:misaligned}%
\end{figure}%

%% file: figures/fig_3.tex
\begin{figure*}%
\def\PicWidth{3.8cm}%
\def\Vgap{1mm}%
\renewcommand{\PlotImage}[1]{\includegraphics[clip, trim=4px 0px 2px 20px, width=\PicWidth]{#1}}
\def\ZoomCenter{(.63,.78)}%
\def\ZoomPos{(.48,.72)}%
\hspace*{5mm}\PlotFig{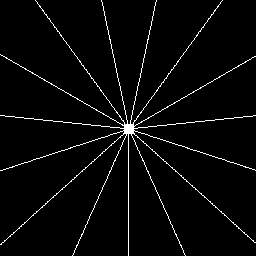}{sampling pattern}\hfill%
\PlotFigZoomSpy{\ZoomCenter}{\ZoomPos}{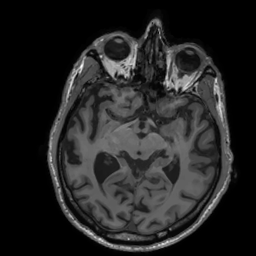}{deformed ground truth}{red}\hfill%
\PlotFigZoomSpy{\ZoomCenter}{\ZoomPos}{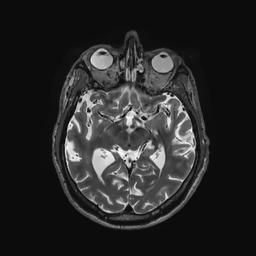}{side information}{red}\hfill%
\PlotFigZoomSpy{\ZoomCenter}{\ZoomPos}{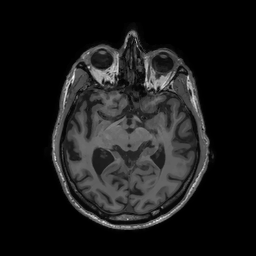}{ground truth}{red}\\[\Vgap]%
\PlotFigZoomSpyMetrics{\ZoomCenter}{\ZoomPos}{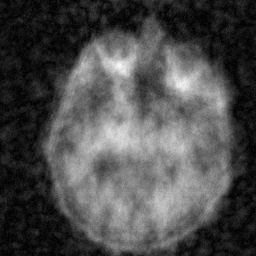}{pseudo-inverse}{36.4\%}{n/a}{red}\hfill%
\PlotFigZoomSpyMetrics{\ZoomCenter}{\ZoomPos}{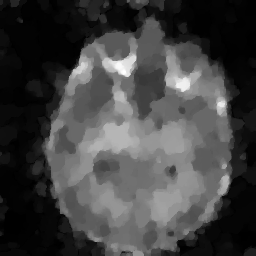}{$\tv$}{56.0\%}{n/a}{red}\hfill%
\renewcommand{\PlotImage}[1]{\includegraphics[clip, trim=4px 0px 2px 15px, width=\PicWidth]{#1}}
\PlotFigZoomSpyMetrics{\ZoomCenter}{\ZoomPos}{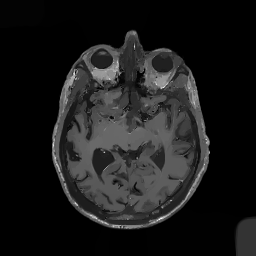}{three-step}{90.4\%}{3.2\%}{red}\hfill%
\PlotFigZoomSpyMetrics{\ZoomCenter}{\ZoomPos}{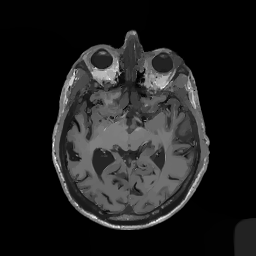}{proposed}{91.7\%}{0.9\%}{red}%
\caption{Multi-contrast MRI reconstruction. Pseudo-inverse, $\tv$, and misaligned $\dtv$ do not correct motion. The proposed method corrects the deformation and the reconstruction satisfyingly agrees with the ground truth. We give (SSIM, RD) values of the reconstructed images and deformation parameters. For pseudo-inverse and $\tv$ no deformation was corrected and SSIM is given with respect to the deformed ground truth.
\label{fig:MRI_comparison}}%
\end{figure*}

%% file: figures/fig_4.tex
\begin{figure*}[h!]
\vspace*{3mm}
\def\PicWidth{3.6cm}%
\newlength{\VfWidth}
\setlength{\VfWidth}{\PicWidth}
\setlength{\VfWidth}{1.2143\VfWidth}
\def\transparency{0.7}
\renewcommand{\PlotFig}[3]{%
\begin{tikzpicture}[spy using outlines, x=\PicWidth, y=\PicWidth, inner sep=0pt]%
\draw (0, 0) node [anchor=center] {{\PlotImage{#1}}};%
\draw (0, 0) node [anchor=center] {{\transparent{\transparency}\includegraphics[clip, trim=45px 30px 14px 14px, width=\VfWidth,angle=90]{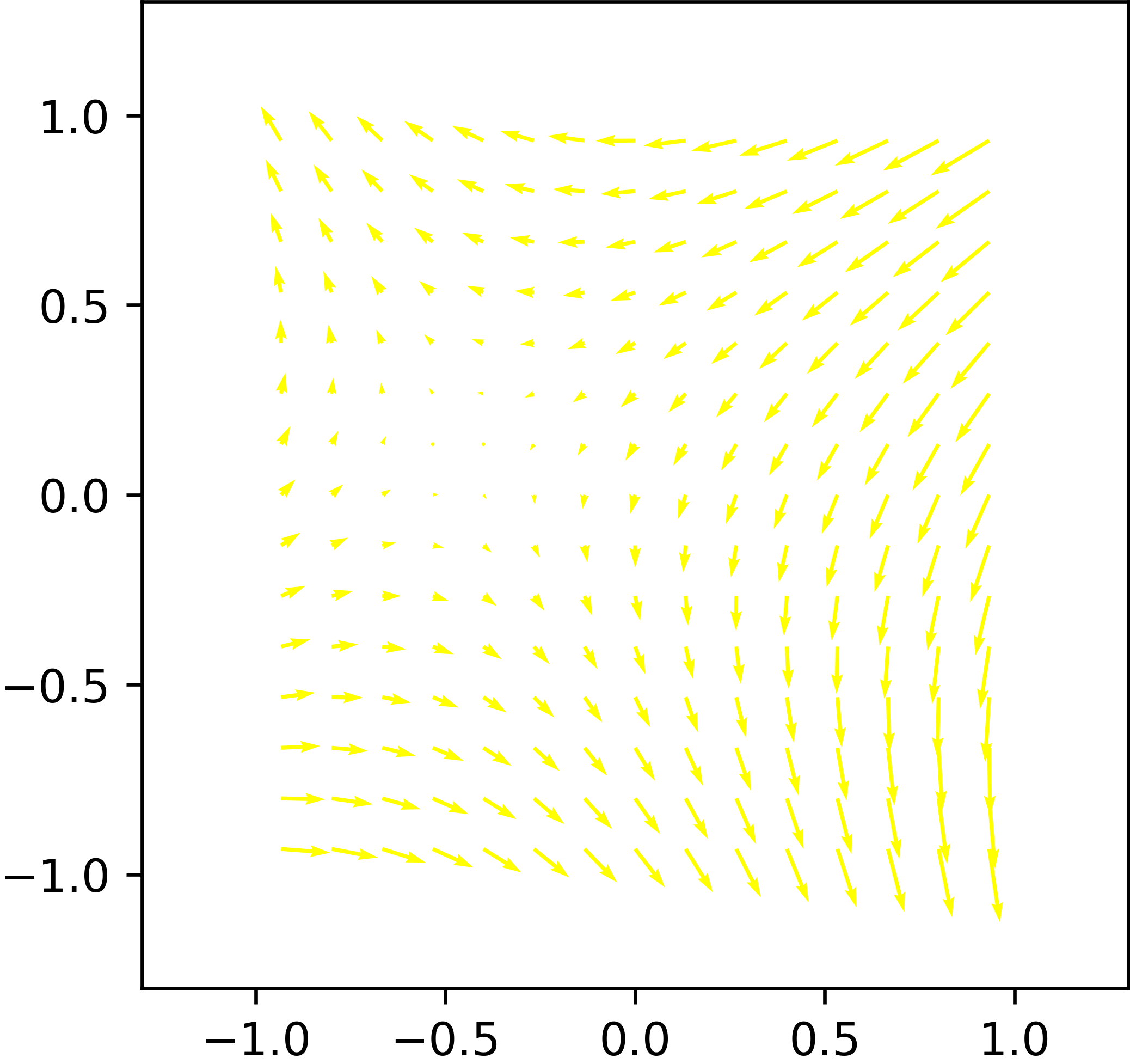}}};%
\draw (0, 0) node [anchor=center] {{\transparent{\transparency}\includegraphics[clip, trim=45px 30px 14px 14px, width=\VfWidth,angle=90]{#2}}};%
\draw (0, -0.55) node [anchor=north] {\color{black}\footnotesize #3\phantom{fg}};%
\end{tikzpicture}}%
\renewcommand{\PlotImage}[1]{{\includegraphics[clip, trim=0px 0px 0px 0px, height=\PicWidth]{#1}}}%
\PlotFig{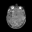}{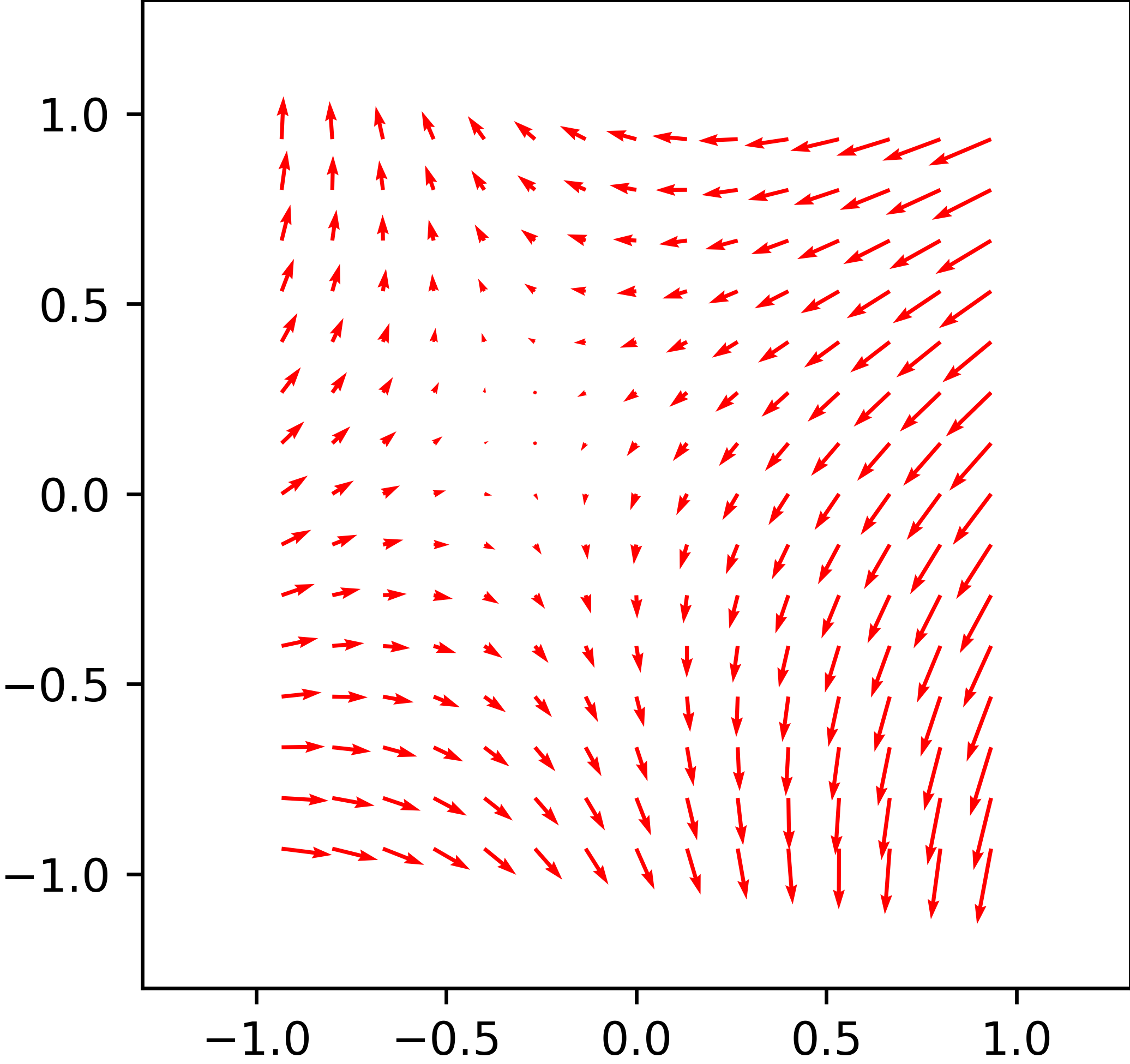}{$32\times 32$, RD 44.1\%}\hfill%
\PlotFig{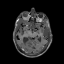}{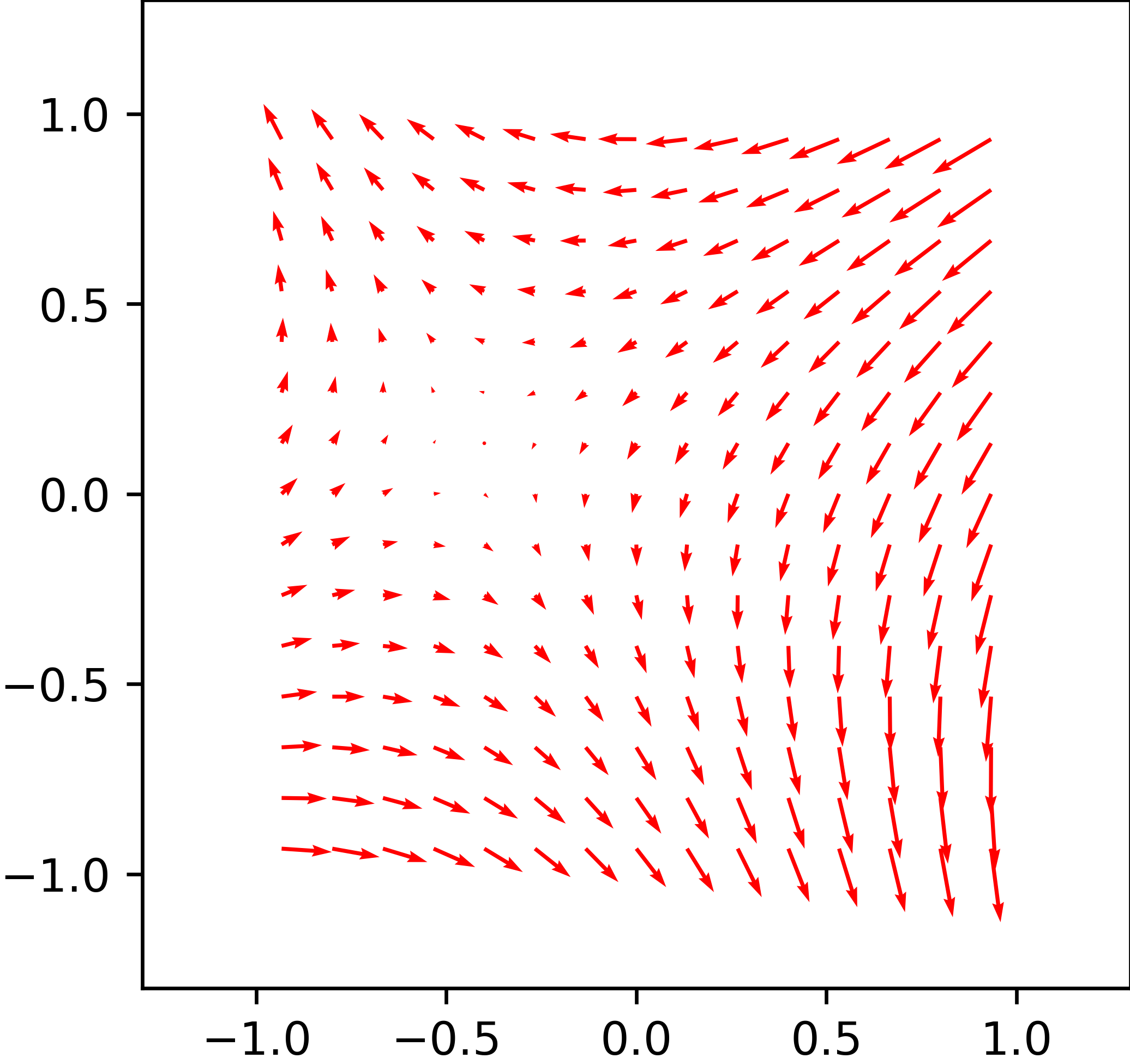}{$64\times 64$, RD 2.8\%}\hfill%
\PlotFig{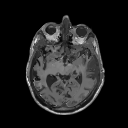}{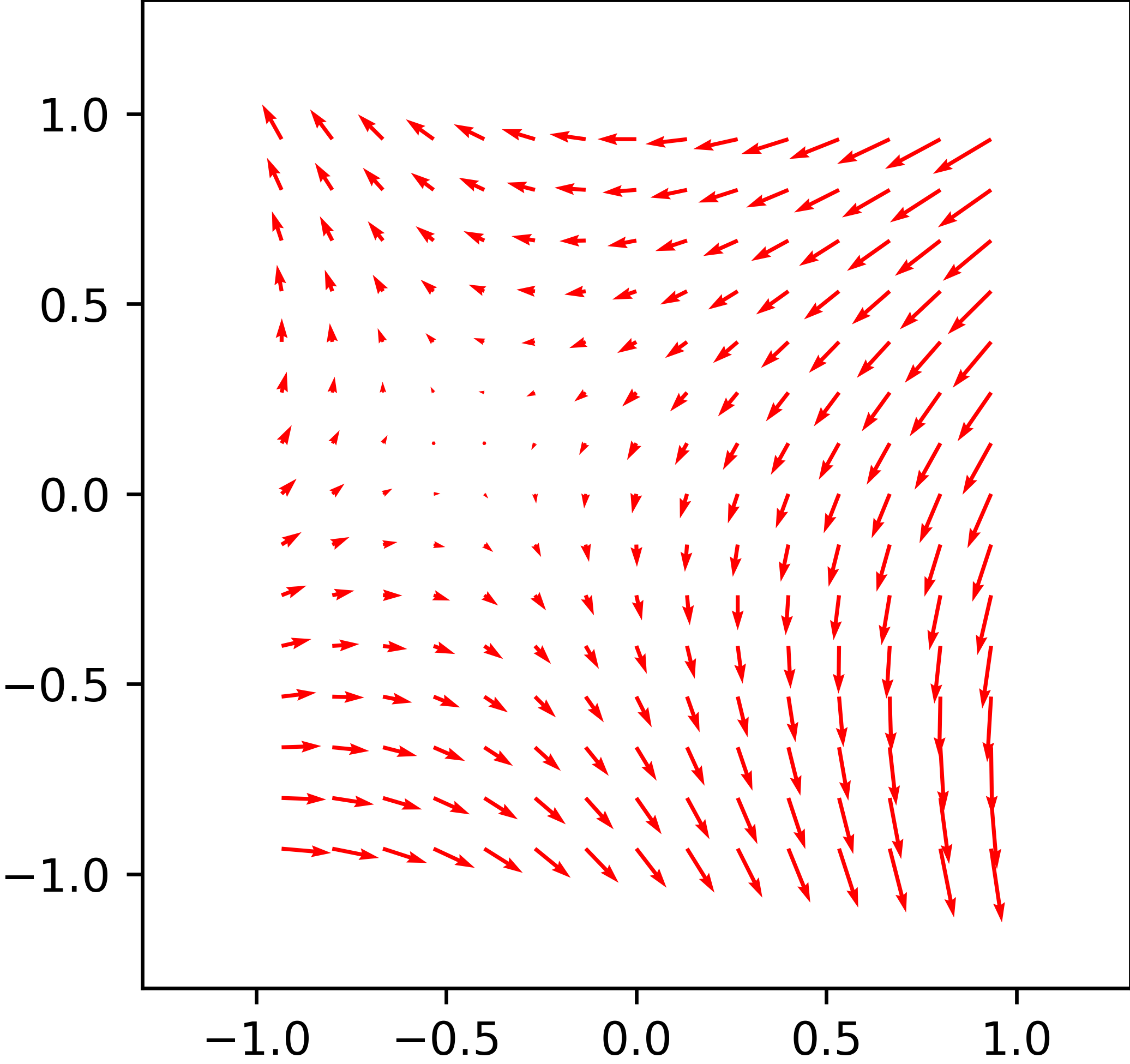}{$128\times 128$, RD 1.3\%}\hfill%
\PlotFig{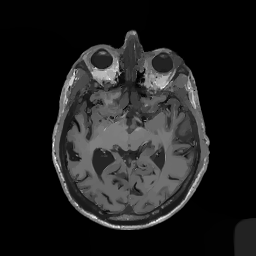}{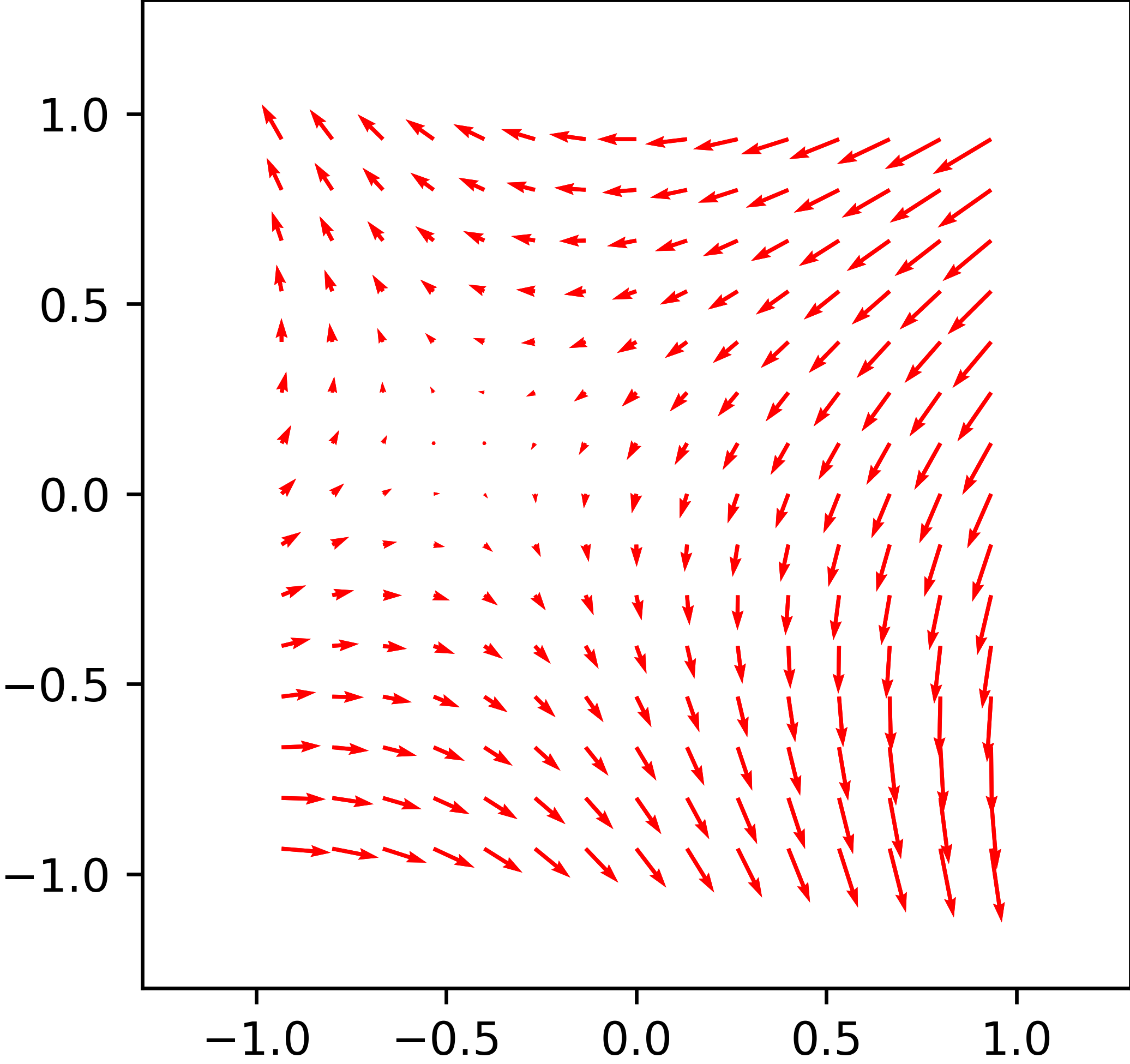}{$256\times 256$, RD 0.9\%}%
\caption{From left to right: MRI reconstructions of increasing resolutions generated by Algorithm~\ref{alg:scale}. The deviation of the ground truth field $\vf_\mathrm{zoom}$ and the reconstructed fields from the identity are shown in yellow and red, respectively. The ground truth field is only visible in the first image where the estimated deformation field is still inaccurate. Relative differences (RD) between reconstructed and ground truth motion parameters are given, as well. \label{fig:MRI_scale_space}}%
\end{figure*}%

%% file: figures/fig_5.tex
\begin{figure*}[h!]%
\def\Vgap{1mm}
\renewcommand{\PlotImage}[1]{\includegraphics[width=\PicWidth, clip, trim=4pt 5pt 2pt 20pt]{#1}}%
\centering%
\def\ZoomCenter{(.65,.3)}%
\def\ZoomPos{(.48,.64)}%
\hspace*{8.5mm}%
\def\PicWidth{3.4cm}%
\PlotFig{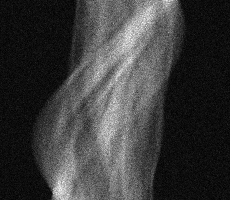}{sinogram data} \hfill%
\def\PicWidth{3.8cm}%
\PlotFigZoomSpy{\ZoomCenter}{\ZoomPos}{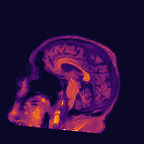}{deformed ground truth}{yellow} \hfill%
\PlotFigZoomSpy{\ZoomCenter}{\ZoomPos}{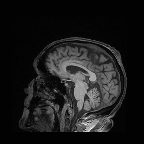}{side information}{yellow}\hfill%
\PlotFigZoomSpy{\ZoomCenter}{\ZoomPos}{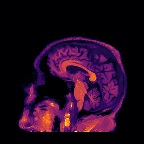}{ground truth}{yellow}\\[\Vgap]%
\PlotFigZoomSpyMetrics{\ZoomCenter}{\ZoomPos}{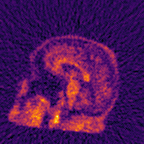}{FBP}{61.3\%}{n/a}{yellow} \hfill%
\PlotFigZoomSpyMetrics{\ZoomCenter}{\ZoomPos}{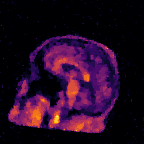}{$\tv$}{88.1\%}{n/a}{yellow}\hfill%
\PlotFigZoomSpyMetrics{\ZoomCenter}{\ZoomPos}{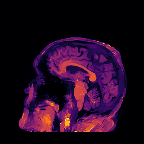}{three-step}{93.3\%}{6.5\%}{yellow}\hfill
\PlotFigZoomSpyMetrics{\ZoomCenter}{\ZoomPos}{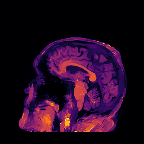}{proposed}{93.6\%}{3.1\%}{yellow}
\caption{PET reconstructions with structural MR side information. Filtered back-projection (FBP) and $\tv$ do not correct motion and yield poor reconstructions. Both the three-step and the proposed method correct the deformation and the reconstruction satisfyingly agrees with the ground truth. We give SSIM and RD values of the reconstructed images and deformation parameters. For FBP and $\tv$ no deformation was corrected and SSIM is given with respect to the deformed ground truth.\label{fig:PET_comparison}}%
\end{figure*}

%% file: figures/fig_6.tex
\begin{figure*}[t]
\vspace*{3mm}
\def\PicWidth{2.8cm}%
\setlength{\VfWidth}{\PicWidth}
\setlength{\VfWidth}{1.2143\VfWidth}
\def\transparency{0.7}
{%
\renewcommand{\PlotFig}[3]{%
\begin{tikzpicture}[spy using outlines, x=\PicWidth, y=\PicWidth, inner sep=0pt]%
\draw (0, 0) node [anchor=center] {{\PlotImage{#1}}};%
\draw (0, 0) node [anchor=center] {{\transparent{\transparency}\includegraphics[clip, trim=45px 30px 14px 14px, width=\VfWidth,angle=90]{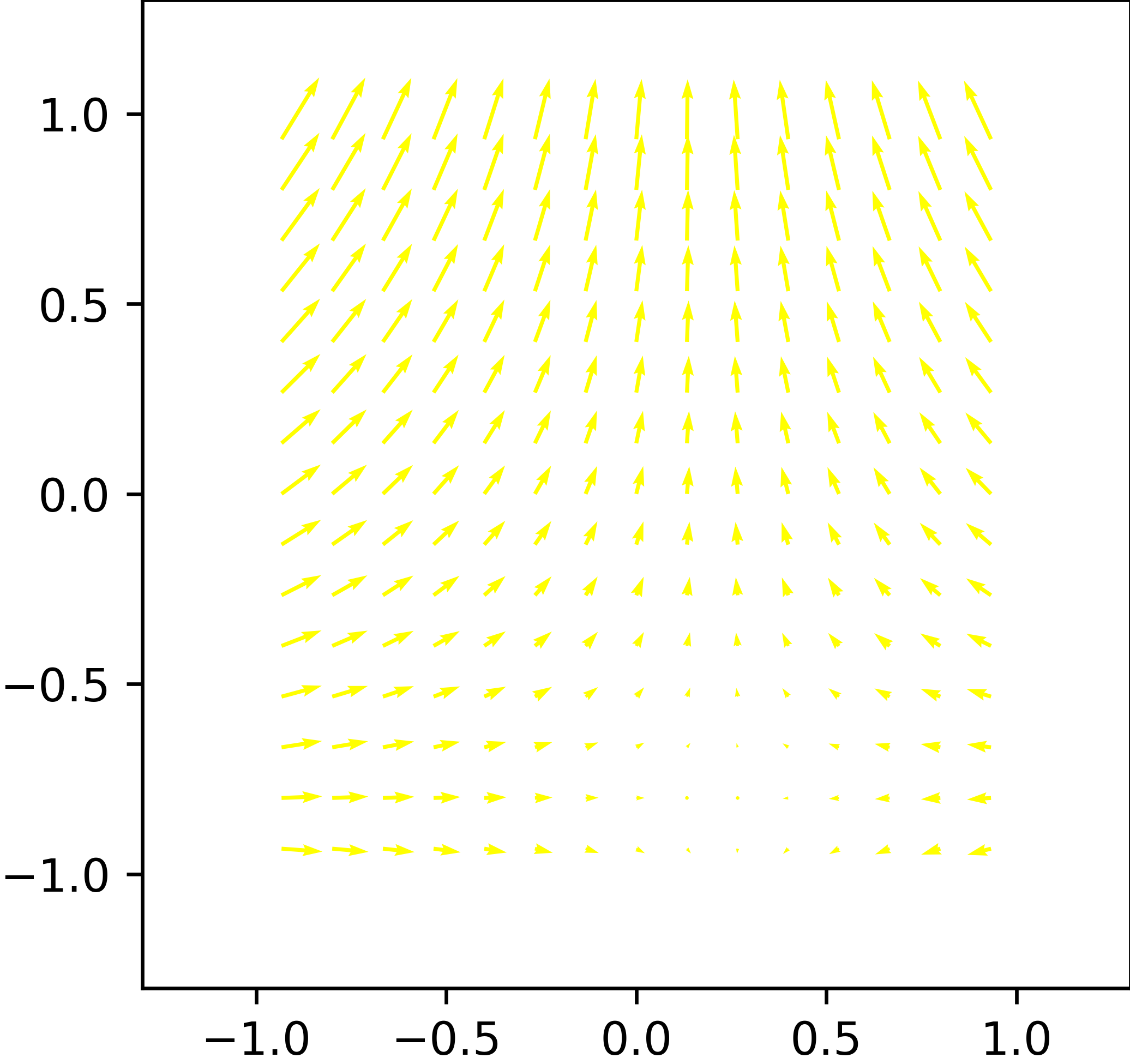}}};%
\draw (0, 0) node [anchor=center] {{\transparent{\transparency}\includegraphics[clip, trim=45px 30px 14px 14px, width=\VfWidth,angle=90]{#2}}};%
\draw (0, -0.55) node [anchor=north] {\color{black}\footnotesize #3\phantom{fg}};%
\end{tikzpicture}}%
\renewcommand{\PlotImage}[1]{{\includegraphics[clip, trim=0px 0px 0px 0px, height=\PicWidth]{#1}}}%
\PlotFig{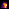}{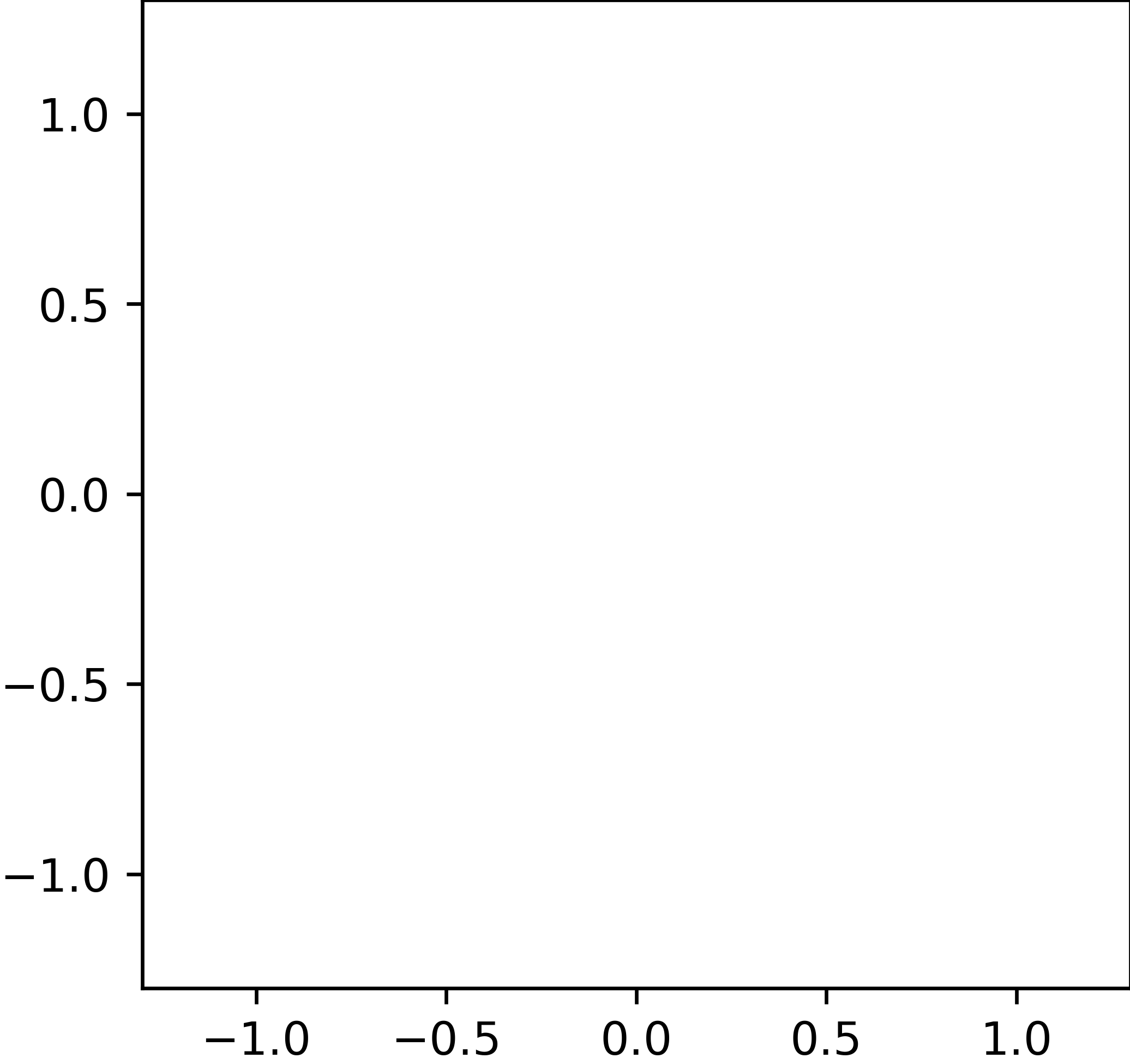}{$9\times 9$, RD 100.0\%}\hfill%
\PlotFig{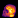}{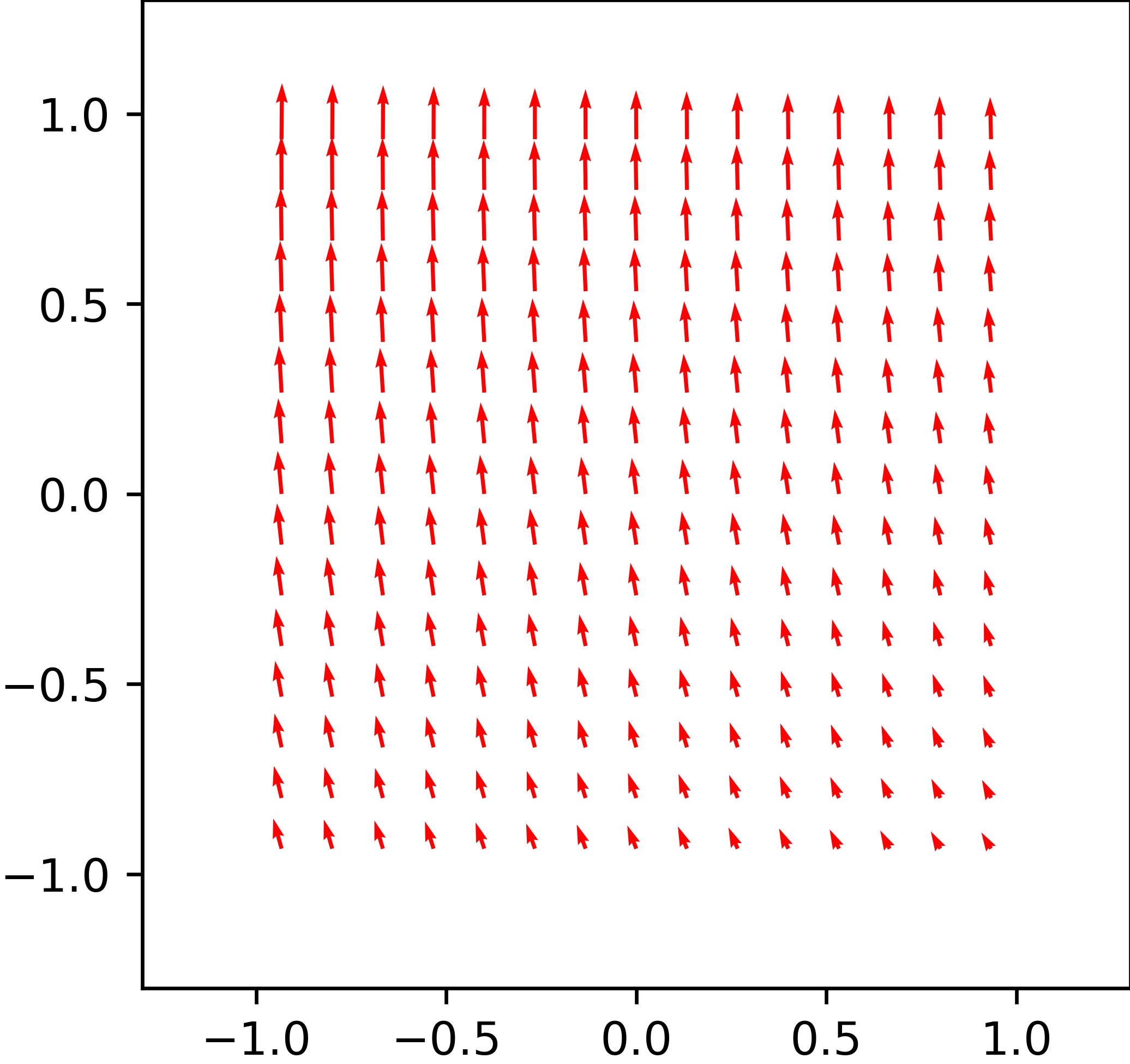}{$18\times 18$, RD 81.7\%}\hfill%
\PlotFig{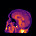}{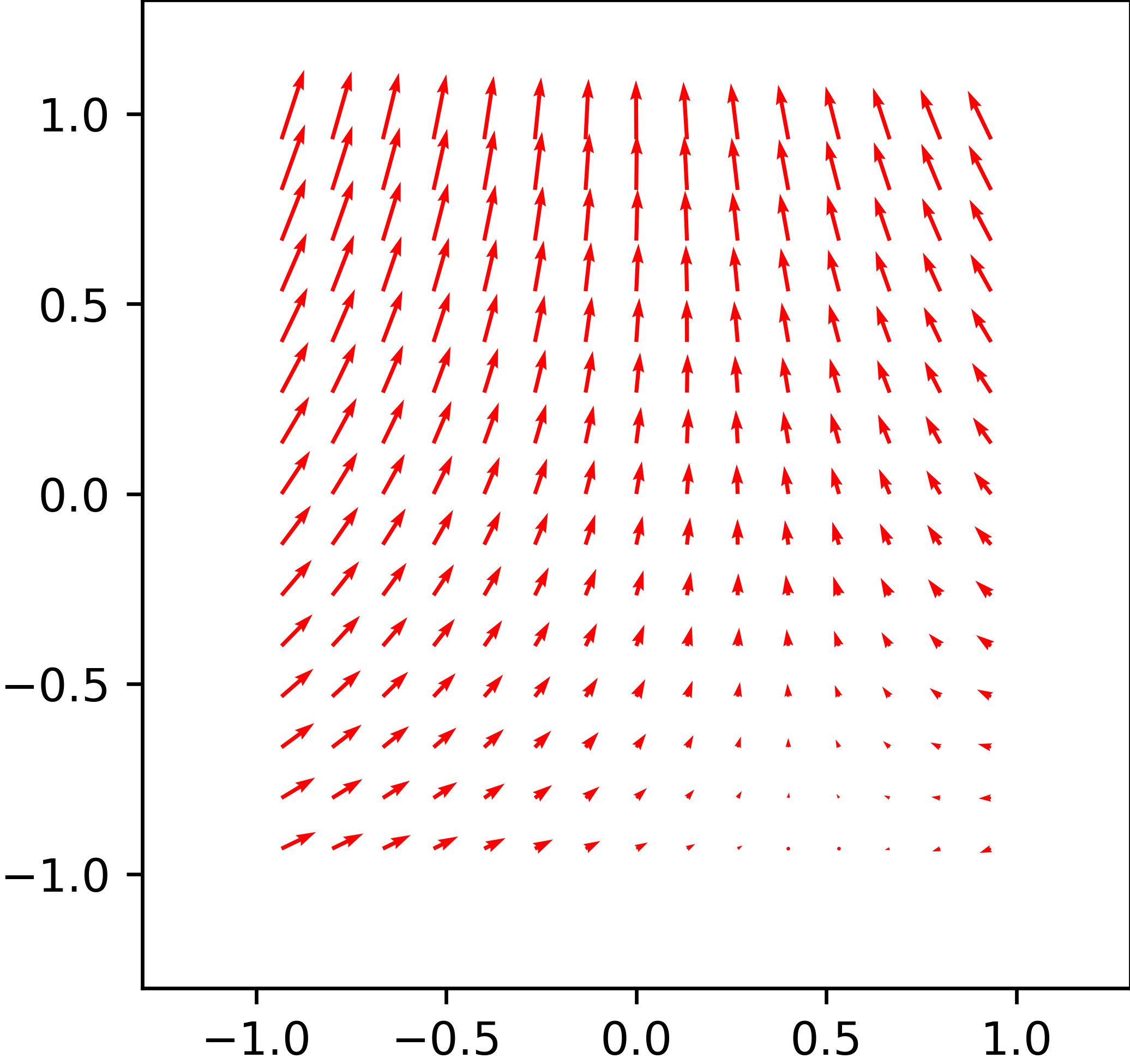}{$36\times 36$, RD 35.5\%}\hfill%
\PlotFig{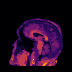}{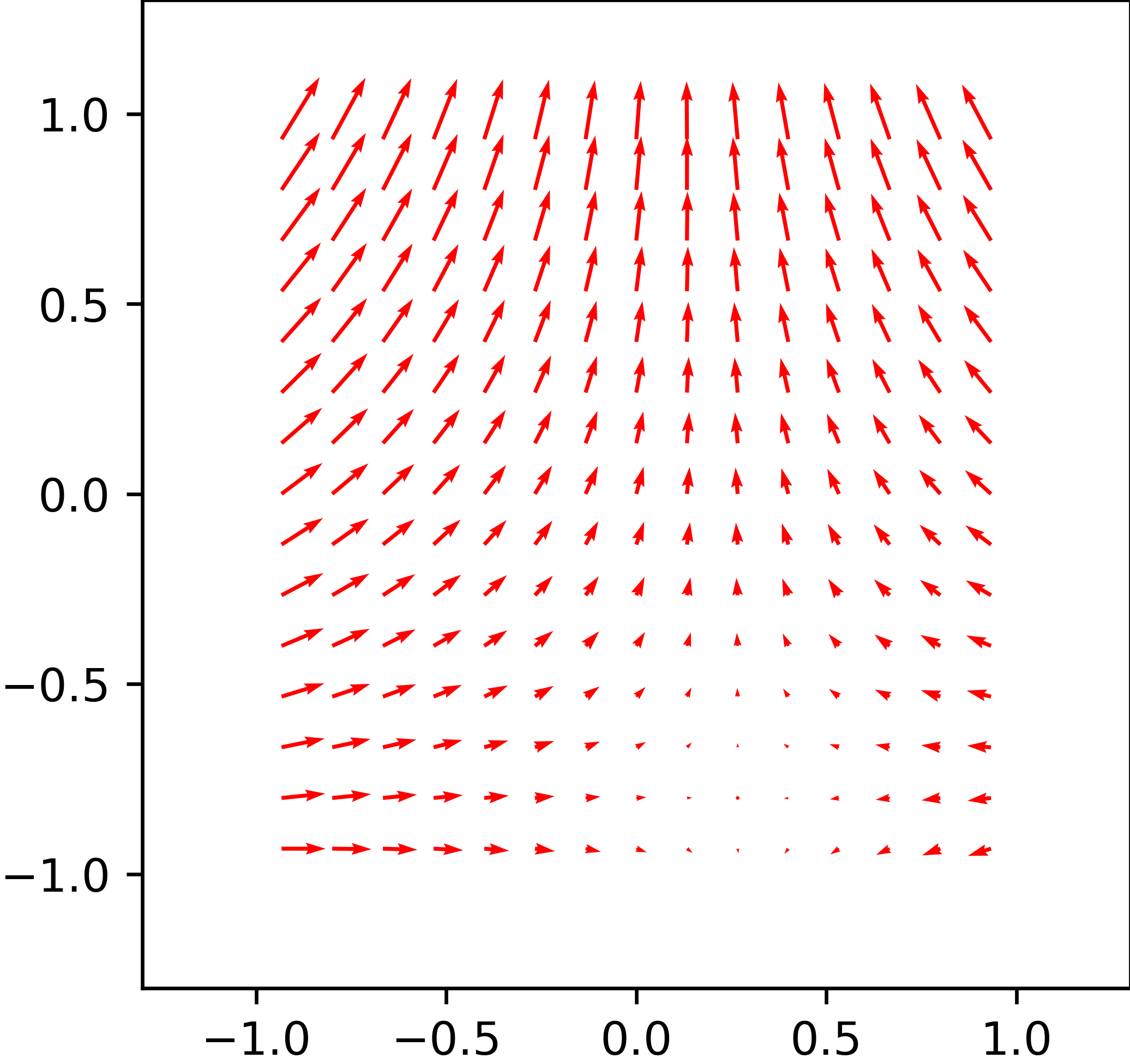}{$72\times 72$, RD 8.0\%}%
\hfill
\PlotFig{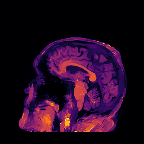}{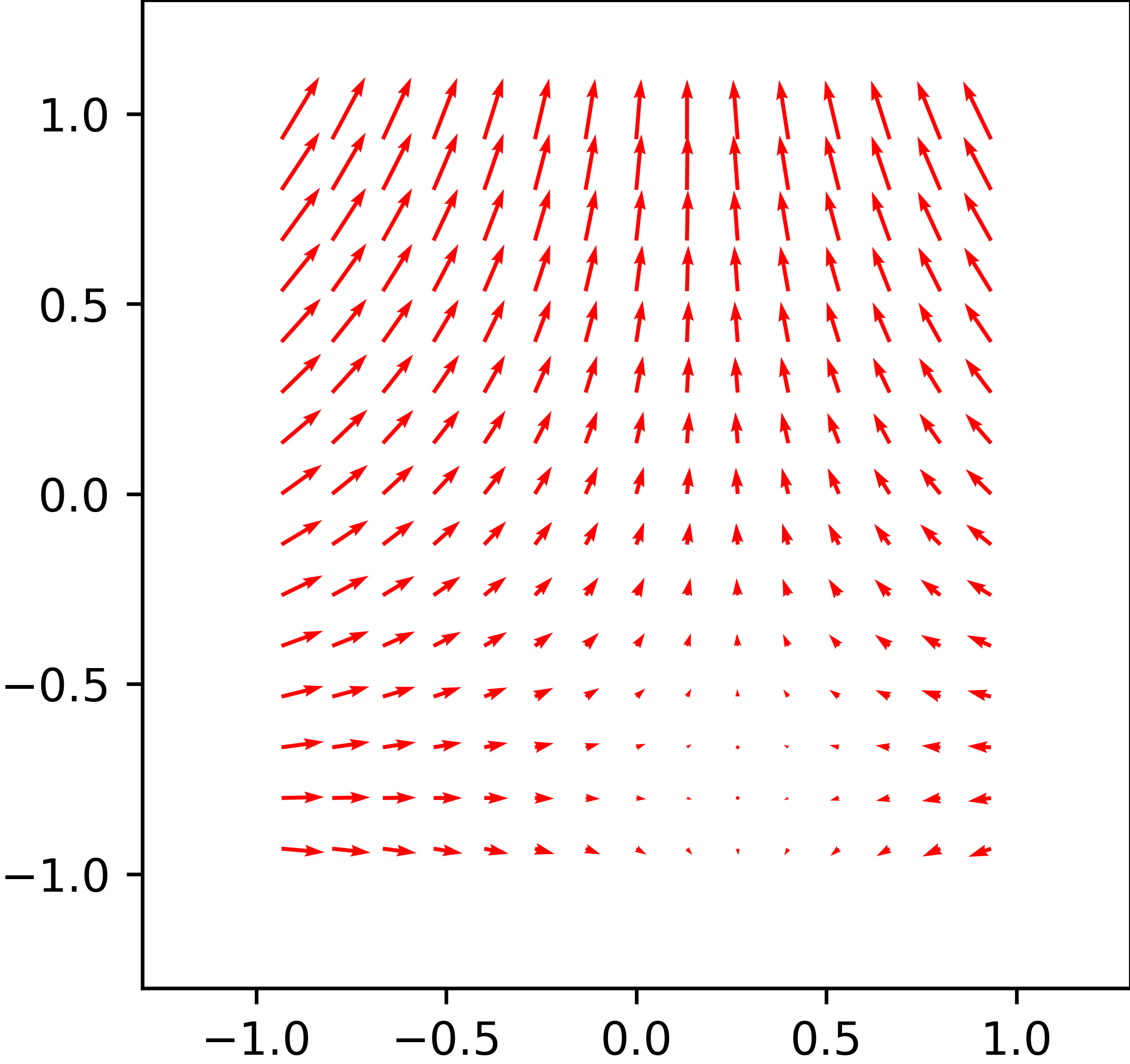}{$144\times 144$, RD 3.1\%}
}
\caption{From left to right: PET reconstructions of increasing resolutions generated by Algorithm~\ref{alg:scale}. The deviation of the ground truth field $\vf_\rigid$ and the reconstructed fields from the identity are shown in yellow and red, respectively. The ground truth field is only visible in the first three images where the estimated deformation field is still inaccurate. In the first image the reconstructed field is zero and thus not plotted. Relative differences (RD) between reconstructed and ground truth motion parameters are given, as well.
\label{fig:PET_scale_space}}%
\end{figure*}
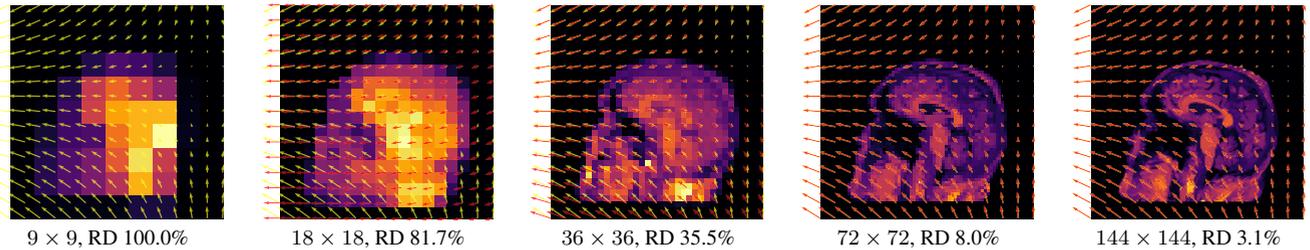

%% file: figures/fig_7.tex
\begin{figure*}
\def\PicWidth{2.57cm}
\renewcommand{\PlotImage}[1]{\includegraphics[cframe=black 2pt,clip, trim=12px 10px 12px 20px, width=\PicWidth]{#1}}%
\newcommand{\PlotRedImage}[1]{\includegraphics[cframe=red 2pt,clip, trim=12px 10px 12px 20px, width=\PicWidth]{#1}}%
\newcommand{\PlotFigMetric}[3]{%
\begin{tikzpicture}[x=\PicWidth, y=\PicWidth, inner sep=0pt]%
\draw (0, 0) node [anchor=south west] {{\PlotImage{#1}}};%
\draw (0.35, 0.0) node [anchor=south west] {\parbox{1.5cm}{\color{white}\footnotesize 
\begin{tabular}{lr}
     SSIM & #2\\
     RD & #3
\end{tabular}
}};%
\end{tikzpicture}}%
\newcommand{\PlotRedFigMetric}[3]{%
\begin{tikzpicture}[x=\PicWidth, y=\PicWidth, inner sep=0pt]%
\draw (0, 0) node [anchor=south west] {{\PlotRedImage{#1}}};%
\draw (0.3, 0.0) node [anchor=south west] {\parbox{1.5cm}{\color{white}\footnotesize 
\begin{tabular}{lr}
     SSIM & #2\\
     RD & #3
\end{tabular}
}};%
\end{tikzpicture}}%
\begingroup
\setlength{\tabcolsep}{1pt}
\begin{tabular}{ccccccc}
    &
   \PlotFig{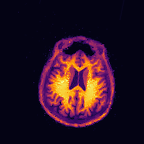}{$11.5^\circ$}  &  
   \PlotFig{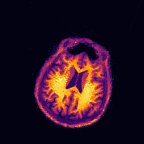}{$22.9^\circ$} &
    \PlotFig{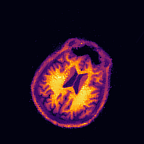}{$34.4^\circ$} &
    \PlotFig{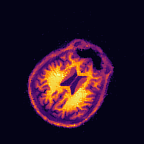}{$45.8^\circ$} &
    \PlotFig{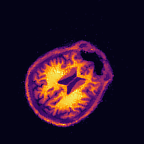}{$57.3^\circ$} &
    \PlotFig{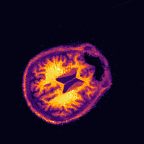}{$68.8^\circ$} \\
    \rotatebox{90}{\quad three-step} &
    \PlotFigMetric{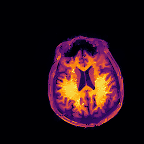}{93.9\%}{7.5\%} &
    \PlotFigMetric{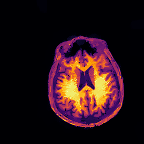}{94.7\%}{2.9\%} &
    \PlotFigMetric{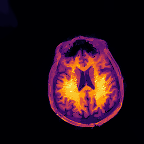}{93.3\%}{2.3\%} &
    \PlotRedFigMetric{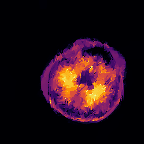}{68.7\%}{76.6\%} &
    & \\
    \rotatebox{90}{\qquad$M=1$} &
    \PlotRedFigMetric{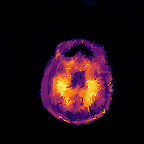}{68.2\%} {100.8\%}\hspace*{\fill} &&&&& \\
    \rotatebox{90}{\qquad$M=2$} &
    \PlotFigMetric{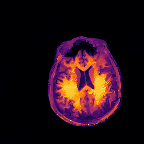}{90.1\%}{14.0\%} &
    \PlotRedFigMetric{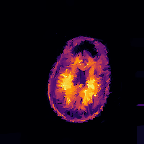}{73.3\%}{84.2\%} &
    &&& \\
    \rotatebox{90}{\qquad$M=3$} & \PlotFigMetric{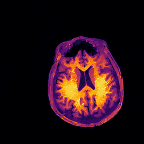}{95.1\%}{4.2\%} &
    \PlotFigMetric{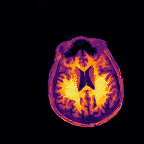}{94.8\%}{4.1\%} &
    \PlotFigMetric{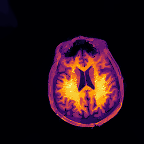}{94.0\%}{2.0\%} &
    \PlotFigMetric{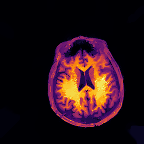}{93.3\%}{1.2\%} &
    \PlotRedFigMetric{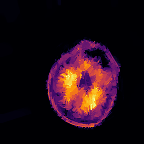}{74.9\%}{68.7\%} \\
    \rotatebox{90}{\qquad$M=4$} &
    \PlotFigMetric{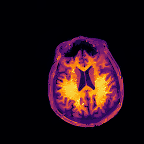}{95.1\%}{3.8\%} &
    \PlotFigMetric{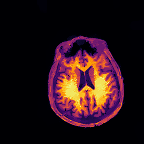}{95.0\%}{3.6\%} &
    \PlotFigMetric{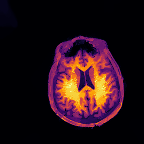}{94.2\%}{1.8\%} &
    \PlotFigMetric{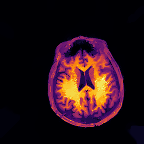}{93.3\%}{1.2\%} &
    \PlotRedFigMetric{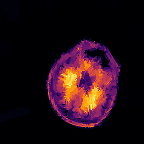}{74.2\%}{69.7\%} &\\
    \rotatebox{90}{\qquad$M=5$} &
    \PlotFigMetric{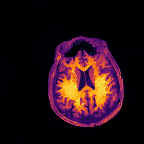}{95.1\%}{4.1\%} &
    \PlotFigMetric{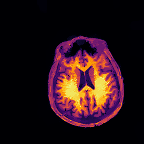}{95.0\%}{3.5\%} &
    \PlotFigMetric{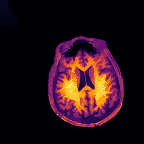}{94.1\%}{1.7\%} &
    \PlotFigMetric{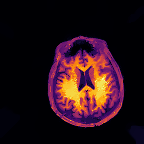}{93.3\%}{1.3\%} &
    \PlotFigMetric{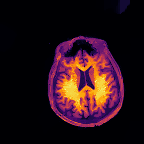}{94.6\%}{1.0\%} &
    \PlotRedFigMetric{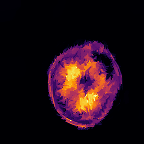}{74.7\%}{84.2\%}
\end{tabular}
\endgroup
\caption{Testing the limits of our method: \textbf{first row:} rotated images which were used to create the PET data, \textbf{second row:} results of the three-step method, \textbf{other rows:} proposed reconstructions for difference sizes $M$ of the scale space. Failed registrations are marked with a red frame. Note that relative differences are larger for small displacements since these have a small norm (cf.~\eqref{eq:param_affine} and \eqref{eq:relative_diff}).\label{fig:rotations}}
\end{figure*}
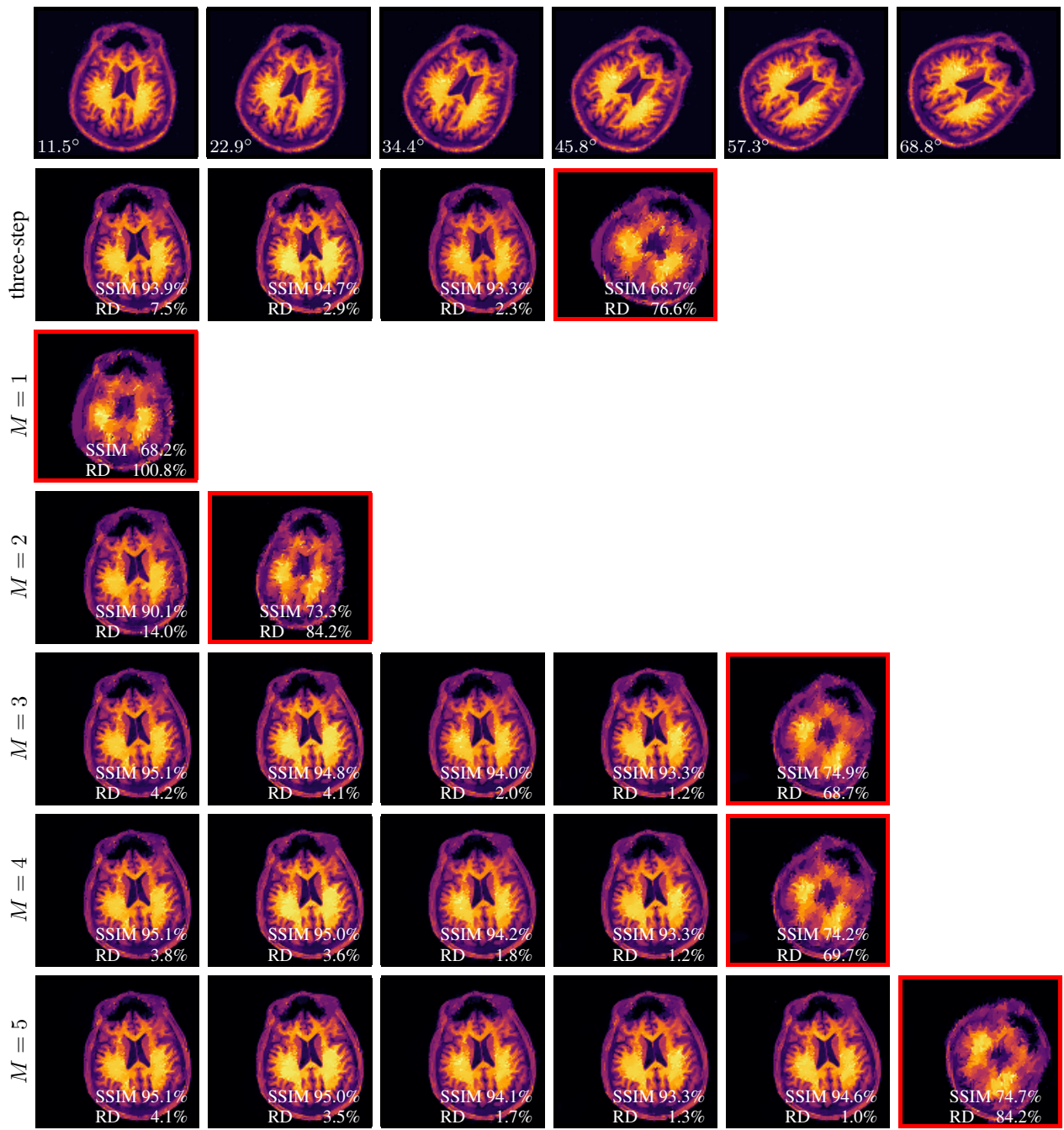%

%% file: figures/fig_8.tex
\begin{figure}%
\def\PicWidth{4cm}%
\centering%
\def\Vgap{1mm}
\def\ZoomCenter{(.68,.7)}
\def\ZoomPos{(0.5, 0.8)}
\PlotFigZoomSpy{\ZoomCenter}{\ZoomPos}{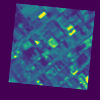}{data \rigid}{red}
\PlotFigZoomSpy{\ZoomCenter}{\ZoomPos}{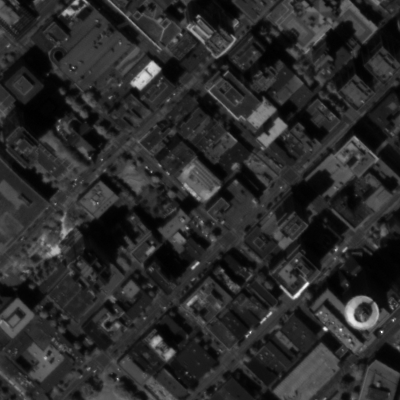}{side information}{red} \\[\Vgap]%
\PlotFigZoomSpy{\ZoomCenter}{\ZoomPos}{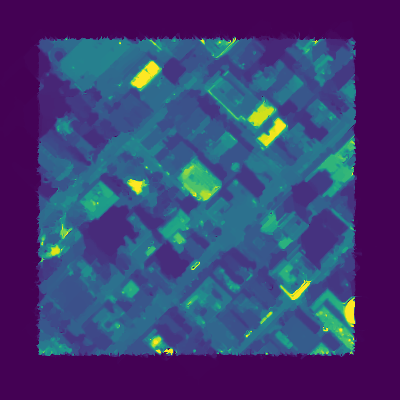}{target}{red}%
\PlotFigZoomSpyMetrics{\ZoomCenter}{\ZoomPos}{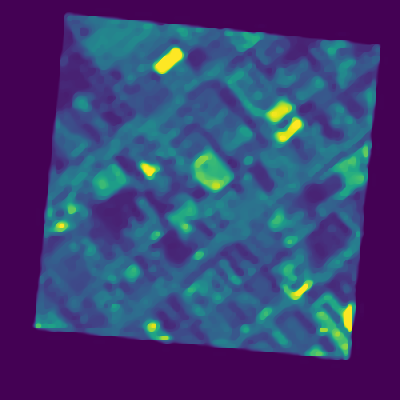}{$\tv$}{89.6\%}{n/a}{red} \\[\Vgap]%
\PlotFigZoomSpyMetrics{\ZoomCenter}{\ZoomPos}{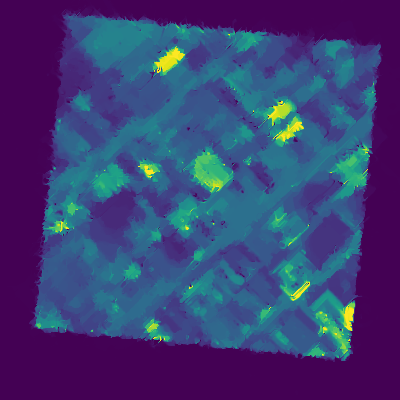}{misaligned $\dtv$}{65.0\%}{n/a}{red}%
\PlotFigZoomSpyMetrics{\ZoomCenter}{\ZoomPos}{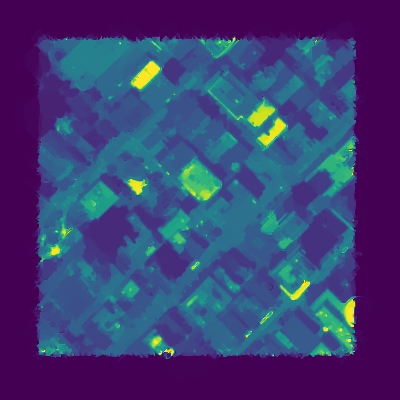}{proposed}{98.2\%}{5.5\%}{red}%
\caption{Hyperspectral super-resolution. Reconstruction of data set \rigid{} using three different methods. $\tv$ and misaligned $\dtv$ do not correct deformations and yield poor results. The proposed method corrects the deformation and the reconstruction is in very good agreement with the target. We give SSIM and RD values of the reconstructed images and deformation parameters. For $\tv$ and misaligned $\dtv$ no deformation was corrected and for $\tv$ SSIM is given with respect to a deformed target image.
\label{fig:RS_comparison}}
\end{figure}

%% file: figures/fig_9.tex
\begin{figure}%
\def\PicWidth{4cm}%
\centering%
\def\Vgap{1mm}
\def\ZoomCenter{(.68,.7)}
\def\ZoomPos{(0.5, 0.81)}
\PlotFigZoomSpy{\ZoomCenter}{\ZoomPos}{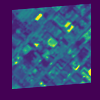}{data \shear}{red}%
\PlotFigZoomSpyMetrics{\ZoomCenter}{\ZoomPos}{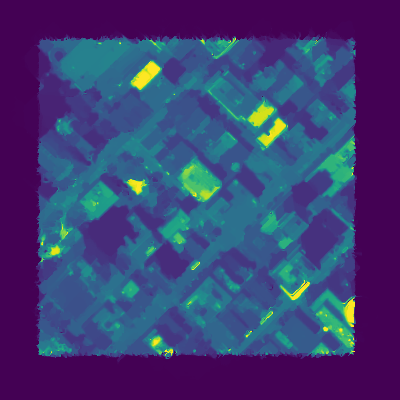}{proposed}{98.9\%}{7.7\%}{red}%
\\[\Vgap]
\PlotFigZoomSpy{\ZoomCenter}{\ZoomPos}{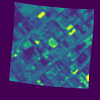}{data \nonlinear}{red}%
\PlotFigZoomSpyMetrics{\ZoomCenter}{\ZoomPos}{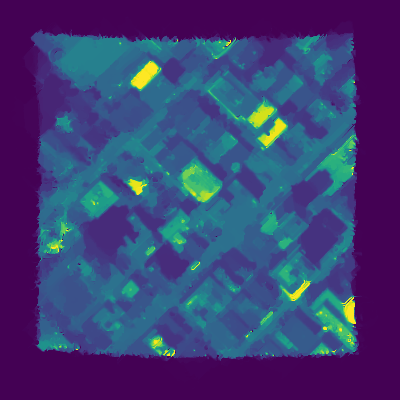}{proposed}{94.3\%}{n/a}{red}%
\caption{Reconstructions of data sets \shear{} and \nonlinear{}.
The proposed method corrects the deformations and yields very good reconstructions. We give SSIM values of the reconstructed images and deformation parameters for both data sets and RD values for the affine data set \shear{}.
\label{fig:RS_recons}}
\end{figure}